\documentclass[useAMS,usenatbib]{mn2e}
\usepackage{aas_macros}
\usepackage[dvips]{graphicx}
\usepackage{amssymb,amsmath}
\usepackage{times}

\title[Tidal Effects \& Environment Dependence of Halo Assembly]
{Tidal Effects and the Environment Dependence of Halo Assembly}

\author[O. Hahn et al.]{Oliver Hahn$^{1}$\thanks{E-mail: hahn@phys.ethz.ch},
 Cristiano Porciani$^{1,2}$, Avishai Dekel$^{3}$ and C. Marcella Carollo$^{1}$\\
$^{1}$Department of Physics, ETH Zurich, 8093 Zurich, Switzerland\\
$^{2}$Argelander Institut f\"ur Astronomie, Auf dem H\"ugel 71, 53121 Bonn, Germany\\
$^{3}$Racah Institute of Physics, The Hebrew University, Jerusalem 91904, Israel }

\begin{document}

\date{MNRAS in press.}
\pagerange{\pageref{firstpage}--\pageref{lastpage}} \pubyear{2009}
\maketitle

\label{firstpage}
\begin{abstract}
We explore a possible origin for the puzzling anti-correlation between the formation
epoch of galactic dark-matter haloes and their environment density. This
correlation has been revealed from cosmological $N$-body simulations and is in
conflict with the Extended Press-Schechter model of halo clustering. Using similar
simulations, we first quantify the straightforward association of an early
formation epoch with a reduced mass growth rate at late times. We then find 
that a primary driver of suppressed growth, by accretion and mergers, is tidal
effects dominated by a neighbouring massive halo. The tidal effects range from a
slowdown of the assembly of haloes due to the shear along the large-scale
filaments that feed the massive halo to actual mass loss in haloes that pass
through the massive halo. Using the restricted three-body problem, we show 
that haloes are prone to tidal mass loss within 1.5 virial radii of a larger 
halo. Our results suggest that the dependence of formation epoch on environment 
density is a secondary effect induced by the enhanced density of haloes in 
filaments near massive haloes where the tides are strong. Our measures of assembly 
rate are particularly correlated with the tidal field at high redshifts $z\sim1$.
\end{abstract}


%%%%%% SECTION: INTRODUCTION %%%%%%%%%%%%%%%%%%%%%%%%%%%%%%%%%%%%%%%%%%%%%%%%%%%%
\section{Introduction}
Dark-matter haloes provide the gravitational potential wells in which galaxies
form, and as such, the understanding of their assembly process during the
cosmological history is a key element in the theory of galaxy formation.
Focusing on gravitational physics while avoiding the complications associated
with baryonic physics, the study of the growth of dark-matter haloes is the
straightforward, more solid part of the theory of galaxy formation. The process
of bottom-up assembly of dark haloes is described successfully by an approximate
formalism, the Extended Press-Schechter (EPS) or excursion-set model, which
combines the statistics of an initial Gaussian random fluctuation field with the
gravitational instability theory of linear fluctuation growth and with the
non-linear spherical collapse model \citep{PressSchechter1974,BCEK91,Lacey1993}.
In the simplest version of this model, the assembly history of a halo is
predicted to be fully determined by its mass, independently of its environment
\citep[e.g.][]{Bardeen1986,Efstathiou1988,MoWhite1996}. More sophisticated
versions of EPS do involve a certain environment dependence, but they do not
recover the correlation seen in the simulations \citep{Catelan1998,Taruya2000}.
The mere mass dependence, motivated by the simplest EPS model, lies at the basis
of the widely used scheme for galaxy clustering that utilises the Halo
Occupation Distribution (HOD) \citep[see][and references therein]{Cooray2002}.

However, numerical simulations of cosmological structure formation revealed that
the halo assembly history does depend on the density of haloes in its
environment
\citep{Sheth2004,Gao2005,Wechsler05,Harker2006,Maulbetsch06,Hahn06,Gao06,
Croton2007,Jing2007,Wetzel2007a,Neistein2008,Angulo2008,Li2008}. For example,
when the halo history is parameterised by a formation redshift $z_{\rm form}$
corresponding to the time by which the halo had assembled one half of its
current mass, haloes that formed earlier are found to be more strongly clustered
and, equivalently, reside in denser environments. This effect is valid for
haloes that are much smaller than the current characteristic gravitationally
collapsing mass-scale $M_{*}$ ($\sim 10^{13}M_\odot$ at redshift zero) and
reverses for haloes above $M_\ast$. The origin of this and related deviations
from the predictions of EPS, sometimes termed ``assembly bias" \citep{Gao06}, is
not obvious --- it poses a very interesting open theoretical question which
deserves a simple understanding. 

There have been several attempts to understand the assembly bias.
\cite{Sandvik2007} noted that the assembly bias can be incorporated in an
excursion-set scenario where the spherical collapse model is replaced by
ellipsoidal collapse and the density threshold for halo collapse depends on the
shape and size of the local environment. Despite recovering some
environment dependence, the effect found by these authors
is too weak to explain the $N$-body results. However, \cite{Sandvik2007}
considered only the tidal field generated by the linear (Gaussian) 
density perturbations, as derived by \cite{Doroshkevich1970}. $N$-body 
simulations indicate that at late times and small scales, the distributions 
of tidal field eigenvalues depart significantly from those for a Gaussian 
random field \citep[see e.g.][]{Hahn06}. Environmental effects in the 
ellipsoidal collapse model have also been discussed by \cite{Desjacques2007} 
but the model investigated by this author could not reproduce the correct 
sign of the assembly bias at low masses.

\cite{wang06} found that small old haloes tend to reside next to very massive 
haloes. Considering the relative velocity of the dark-matter flow surrounding 
the haloes, they observed that particles surrounding the haloes with highest 
$z_{\rm form}$ have large relative velocities.

Most recently, \cite{Dalal2008} showed that the assembly bias at high masses
follows naturally from the statistics of peaks of Gaussian random fluctuations
when accounting also for the peak curvature.
At low masses, these authors find a systematic trend of halo age with the
velocity dispersion of the environment. Both, \cite{wang06} and \cite{Dalal2008}
suggest that the accretion onto a halo is suppressed because environmental
velocities exceed the halo's virial velocity. A possible scenario leading to
these high velocities is that the dark-matter fluid is ``hotter'' in regions of
higher density \citep[see also][]{Mo2005} after shell-crossing 
during the formation of the host large-scale structure. This is difficult to 
reconcile with other findings that the dark-matter flow in dense filaments is 
rather ``cold" \citep[e.g.][]{Klypin2003}. In particular, \cite{Sandvik2007} 
argue that the mass of the large-scale environment in which the low-mass haloes 
form is not large enough to account for this explanation. Furthermore, the 
measures employed by \cite{wang06} and \cite{Dalal2008} are not sensitive to 
distinguish between various scenarios: ``hot'' flows, haloes that move against 
the flow of their environment on ``unorthodox'' orbits \citep{Ludlow2008} and 
strongly sheared flows.

In this article, we address the hypothesis that the early formation times of
haloes of a given mass much below the non-linear mass in dense environments are
primarily driven by the {\em tidal} suppression of the halo growth rate in the
vicinity of a neighbouring massive halo. Small, typical (low-sigma) haloes tend
to form in filaments \citep[e.g.][]{Porciani02}, and stream along them into more
massive, rare (high-sigma) haloes. The tidal field in these regions is likely to
generate an ordered shear flow and thus suppress the halo growth by 
accretion or mergers. This
suppression is expected to be correlated with the proximity to the massive halo.
Those small haloes that have already passed through the inner regions of bigger
haloes are likely to have lost mass due to tidal stripping
\citep{Gill2005,Diemand2007a,Ludlow2008} and thus show the earliest formation
times.

In order to test this hypothesis, we quantify in several different ways the halo
assembly history and its environment, and study the correlations between the
adopted quantities. The traditional parameterisation of assembly history via the
formation epoch $z_{\rm form}$ is complemented with more direct measures of halo
growth rate at different redshifts. The environment, commonly referred to by the
halo number density or the amplitude of the two-point correlation function, is
characterised instead by the tidal/shear field, via the eigenvalues of the
deformation tensor. The correlations between assembly history and environment
are shown graphically and measured via correlation coefficients. The primary
role of the tidal field over the density field is to be addressed by the
relative strength of the corresponding correlations with the assembly history.

The outline of this article is as follows. In Section 2, we summarise the
specifics of our $N$-body simulations and the dark-matter halo catalogues. In
Section 3, we recover the dependence of halo formation times on environment
density and clustering amplitude. In Section 4 we introduce further measures of
halo assembly complementing formation redshift. In Section 5, we define the
relevant characteristics of the sheared flow around haloes and its tidal origin
before we illustrate its relation to mass assembly. In Section 6, we demonstrate
the role of tidal effects in the assembly bias, by quantifying the correlation 
between the tidal field and the assembly rate, and comparing it to the 
correlation with environment density. We discuss our analysis in Section 7 and 
provide our conclusions in Section 8.

%%%%%% SECTION: SIMULATIONS %%%%%%%%%%%%%%%%%%%%%%%%%%%%%%%%%%%%%%%%%%%%%%%%
\section{Numerical simulations}

\subsection{Specifics of the $N$-body simulations}
We use a series of three high-resolution cosmological N-body simulations that
were obtained with the tree-PM code {\sc Gadget-2} \citep{Springel2005}. These
simulations are used to follow the non-linear evolution of density perturbations
in a flat $\Lambda$CDM cosmology with a matter density parameter $\Omega_{\rm
m}=0.25$, baryonic contribution $\Omega_{\rm b}=0.045$, and a present-day value
of the Hubble constant $H_0=100\,h\,{\rm km}\,{\rm s}^{-1}$ with $h=0.73$. The
initial power spectrum has a long-wave spectral index of $n=1$, and is
normalised so that the rms fluctuation within a sphere of $8\,h^{-1}$Mpc
linearly extrapolated to the present time is $\sigma_8=0.9$.
Each of the three simulations follows $512^3$ collisionless dark matter
particles in periodic boxes of sizes $45$, $90$ and $180\,h^{-1}{\rm Mpc}$,
respectively. Initial conditions were generated using the {\sc Grafic} tool
\citep{Bertschinger2001} at $z\simeq79$, $65$ and $52$ for the three boxes. The
corresponding particle masses are $4.7\times10^{7}$, $3.8\times10^{8}$ and
$3.0\times10^{9}\,h^{-1}{\rm M}_\odot$, respectively. Particle positions and
velocities were saved for 30 time-steps, logarithmically spaced in expansion
parameter $a$ between $z=10$ and $z=0$. Unless otherwise stated, data from the
simulation with the highest mass resolution, the $45\,h^{-1}{\rm Mpc}$ box, is
used for the analysis of the lowest mass haloes. Results have been checked for
resolution and box size effects against the other boxes.

\subsection{Identification of haloes and the properties of their assembly}
\label{sec:haloes}
We identify bound structures using the standard friends-of-friends (FOF)
algorithm \citep{Davis1985} with a linking length equal to 0.2 times the mean
inter-particle distance. Among all these dark-matter haloes, we keep only those
haloes at $z=0$ that fulfil the virialization constraint $|2T/V+1|<0.5$
\citep{Bett06} where $T$ is the kinetic energy and $V$ is the potential energy
of the group of particles. In addition, we also exclude those groups where the
most-bound particle is further than $R_{\rm max}/4$ away from the centre of
mass, where $R_{\rm max}$ is the distance of the furthest FOF particle from the
group's centre of mass.

For each halo at redshift $z=0$ we find its progenitors at any desired redshift
$z>0$ by identifying all the haloes that have particles in common with the final
halo. At each snapshot, the most massive progenitor that contributes at least
half of its particles to the final halo at $z=0$ is identified as the ``main
progenitor''. We define the formation redshift $z_{\rm form}$ of a halo of mass
$M_0$ at $z=0$ as the redshift at which a main progenitor with mass $M \geq
M_0/2$ first appears in the simulation. The ``assembly history'' of a halo, for
our current purpose, is simply the mass-growth history of its main progenitor,
$M(z)$.

In addition, we say that a halo underwent a ``mass-loss" event if between any
successive snapshots along its history it contributed at least 3\% of its mass
to a more massive halo that is not a main progenitor of the same final
halo.\footnote{This is equivalent to a $\sim5$ per cent change in measured
formation redshift (see eq. \ref{eq:change_zform}).}

\section{The Dependence of Formation Time on Environment}

\begin{figure*}
\begin{center}
	\includegraphics[width=0.45\textwidth]{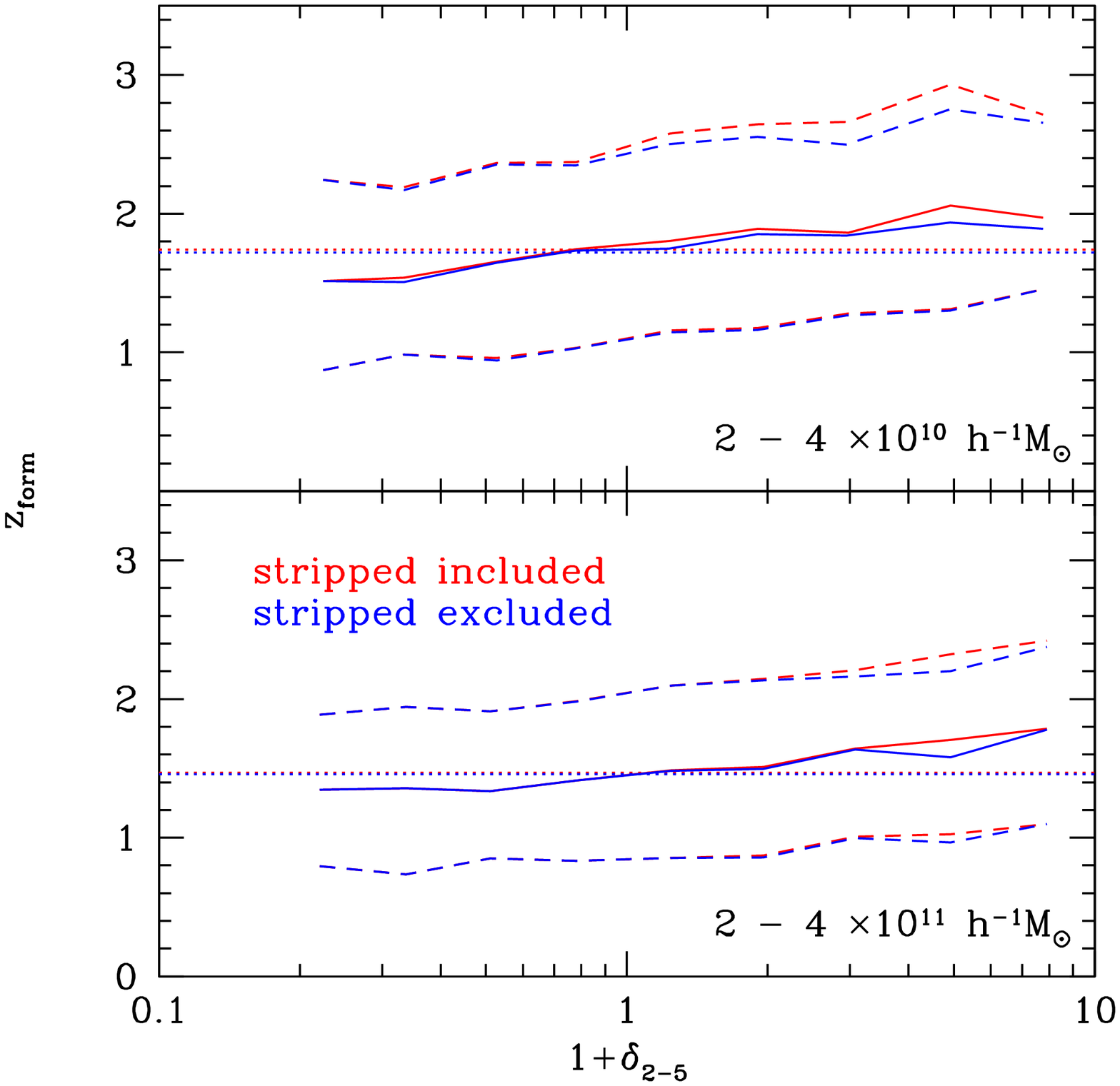}
	\hspace{1.5cm}
	\includegraphics[width=0.45\textwidth]{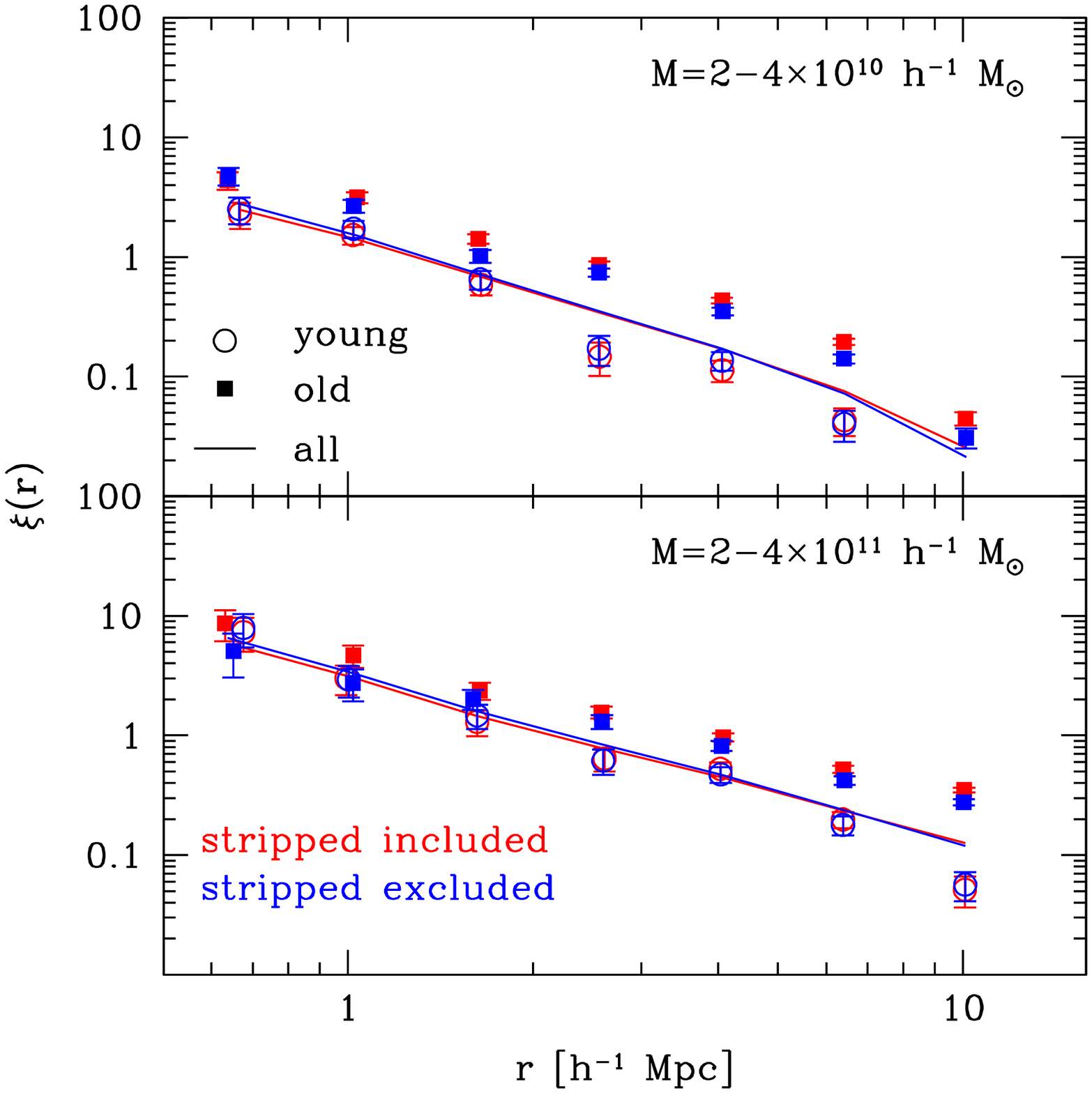}
\end{center}
\caption{\label{fig:harker_massloss}({\it Left Panel}): Median formation
redshift (solid line) as a function of the overdensity $\delta_{2-5}$ (see text
for explanation) for all haloes, including those that experienced tidal mass
loss (red), and only those haloes that never underwent such a mass loss (blue).
The dashed lines indicate the 16th and 84th percentile in each bin. Results are
shown for two different mass-ranges, indicated in the respective panels. The
dotted lines show the median formation redshift for each mass bin. ({\it Right
Panel}): Two-point correlation functions for all haloes (solid lines), and for
those haloes of given mass that have 20 per cent highest/lowest formation times
(filled squares/open circles). Errorbars correspond to Poisson errors.
All results are shown including those haloes that
experience tidal stripping of their mass (red) and excluding them (blue). See
Section \ref{sec:haloes} for the definition of tidally stripped haloes.}
\end{figure*}

The assembly bias has been detected using measures of either environment 
density or halo clustering. We first confirm these findings using our 
simulations.

\subsection{Dependence on Environmental Density}
At a fixed halo mass, the assembly bias can be measured as a systematic trend of
halo formation time with the local density of the environment
\citep[e.g.][]{Harker2006}. We estimate the density at halo positions by
smoothing the distribution of N-body particles on a given scale by convolving
the particle positions with a spherical top-hat kernel of radius $R$. In
practice, we first compute the density on a Cartesian grid using cloud-in-cell
interpolation and then perform the convolution with the kernel using FFT. We
denote with the symbol $\delta_R$ the mass overdensity computed using a
smoothing radius of $R\,h^{-1}$Mpc. Similarly, we call $\delta_{R_1-R_2}$ the
overdensity computed within a spherical shell with an inner radius of
$R_1\,h^{-1}$Mpc and an outer radius of $R_2\,h^{-1}$Mpc.

Figure \ref{fig:harker_massloss} (left panel) shows the median and the 16th and
84th percentiles of $z_{\rm f}$ as a function of $\delta_{2-5}$ for our
simulated haloes in two mass bins. We recover the trend found by
\cite{Harker2006}, and extend it to lower masses. We note that the removal of
mass-loosing haloes from the sample makes a difference preferentially at high
densities and at the high end of the $z_{\rm f}$ distribution. We will return to
these haloes in Section \ref{sec:sim_stripping} where we quantify their
abundance.

\subsection{Dependence on Clustering Amplitude}
Alternatively, the assembly bias can be discussed in terms of the halo
clustering amplitude. The 2-point correlation function  $\xi(r)$ is defined as
the excess probability (with respect to random) to find two haloes in volume
elements ${\rm d}V_1$ and ${\rm d}V_2$ separated by a comoving distance $r$,
\begin{equation}
{\rm d}P_{12}(r) \equiv \bar{n}^2\left[1+\xi(r)\right]\,{\rm d}V_1\,{\rm d}V_2\;,
\end{equation}
where $\bar{n}$ indicates the mean halo number density. The assembly bias has
been originally seen as a systematic dependence of the amplitude of $\xi(r)$
with halo formation time \citep{Gao2005,Gao06}. At constant halo mass, older
haloes tend to be more strongly clustered than younger ones. The age dependence
of the clustering amplitude is evident only for haloes with mass $M<M_\ast(z)$.
Here $M_\ast(z)$ denotes the mass scale for which $1\sigma$ density fluctuations
typically collapse at redshift $z$ according to the Press-Schechter model. The
linear rms overdensity on this mass scale has then to be
$\sigma(M_\ast(z),z)=\delta_c\simeq1.686$. At all redshifts, the number density
of haloes with $M>M_\ast(z)$ is exponentially suppressed.
In Figure \ref{fig:harker_massloss} (right panel) we show how $\xi(r)$ changes
with halo age in our simulations. The ratio of the different correlation
functions agrees very well with the results of \cite{Gao06} and extends the
analysis also to lower masses.

%%%%% SECTION: HALO MASS ASSEMBLY HISTORIES %%%%%%%%%%%%%%%%%%%%%%%%%%%%%%%%%%%%%%%%%
\section{Halo Mass Assembly Histories}
In this section we introduce measures of halo assembly complementing
the common parameterisation of assembly histories through the formation
time.

\subsection{The mass assembly rate of haloes}
\label{sec:assembly_rate}
We define the current assembly rate of each halo by taking the time derivative
of its smoothed assembly history $\mathcal{S}[M(z)]$, where $\mathcal{S}$ is
some smoothing operation. From the smoothed assembly histories, we define the
current assembly rate as \citep[cf.][]{Neistein2006}
\begin{equation}
A(z)\equiv -\frac{\mbox{d}}{\mbox{d}z}\ln\,\mathcal{S}[M(z)].
\end{equation} We apply smoothing by local regression
\citep[LOESS,][]{Cleveland1979}, i.e. the data are fitted by a linear model
inside a window and the fit value is evaluated at the centre of each window
position. A window of $0.5$ in redshift proved to be well suited and has been
used in our analysis. It is necessary to smooth since the assembly histories of
single haloes are prone to fluctuations that are due to the intrinsic
stochasticity of the simulation/halo-identification process.

In close analogy to the definition of the formation redshift, we will use the
ratio $G(z)\equiv M(0)/M(z)$ between the final mass and the mass at redshift $z$
as the ``forward'' measure of halo growth since redshift $z$. 

Both quantities $A(z)$ and $G(z)$ are strongly related with the integral measure
$z_{\rm form}$ (note that $G(z_{\rm form})=2$ by definition of $z_{\rm form}$).
In Table \ref{tab:assemb_corr} we show the Spearman rank correlations between
these different measures as a function of redshift for the main progenitors of
haloes with masses between 2 and $4\times10^{10}$ and 2 and
$4\times10^{11}\,h^{-1}{\rm M}_\odot$ at $z=0$. The median formation redshift
for the first sample of haloes is 1.71 while it is 1.44 for the latter. We use
the rank correlation coefficient rather than the linear Pearson correlation
coefficient as it does not assume a linear relation between the correlated
quantities.

For haloes of a given mass at $z=0$, a large (small) $G(z)$ at high redshift
implies trivially that the main progenitor mass was already high (low) at that
epoch so that its formation time is high (low). The progenitor then also has a
high (low) assembly rate $A(z)$ at early times and a low (high) $A(z)$ at late
times.

Thus, a high formation redshift is related to a low assembly rate $A(z)$ at late
times ($z\lesssim z_{\rm form}$) and a high assembly rate at early times
($z\gtrsim z_{\rm form}$). The latter can be seen for the higher mass sample
where the median formation redshift is sufficiently low so that the correlation
coefficient changes sign at $z=2$. At late times, the formation redshift depends
most sensitively on the assembly rate around $0.5\lesssim z \lesssim 1$ for the
low mass sample and around $z\sim 0.5$ for the higher mass sample.

\begin{table}
\begin{center}
\begin{tabular}{lcccc}
\hline
\multicolumn{1}{c}{$\rho_s$}  & $z=0$ 	& $z=0.5$ 	& $z=1$ 	& $z=2$ \\
\hline
\multicolumn{5}{c}{$M(0)=2-4\times10^{10}\,h^{-1}{\rm M}_\odot$} \vspace{0.2cm}\\
$\left(z_{\rm form},A(z)\right)$ & -0.58 & -0.62 & -0.65 & -0.08 \\
$\left(z_{\rm form},G(z)\right)$  & - & -0.40 & -0.67 & -0.87 \\
$\left(A(z),G(z)\right)$  & - & 0.42 & 0.32 & 0.16 \\
\hline
\hline
\multicolumn{5}{c}{$M(0)=2-4\times10^{11}\,h^{-1}{\rm M}_\odot$} \vspace{0.2cm}\\
$\left(z_{\rm form},A(z)\right)$ & -0.46 & -0.57 & -0.29 & 0.50 \\
$\left(z_{\rm form},G(z)\right)$  & - & -0.40 & -0.77 & -0.81 \\
$\left(A(z),G(z)\right)$  & - & 0.36 & -0.03 & -0.30 \\
\hline
\end{tabular}
\end{center}

\caption{\label{tab:assemb_corr}Spearman rank correlation coefficients $\rho_s$
between three different measures of halo mass assembly: the formation redshift
$z_{\rm form}$, the assembly rate $A(z)$ and the mass growth ratio $G(z)$. The
coefficients are given for the progenitors of haloes identified at $z=0$ with
masses between $2$ and $4\times10^{10}\,h^{-1}{\rm M}_\odot$ (top) in the
$45\,h^{-1}{\rm Mpc}$ box, and for haloes with masses between $2$ and
$4\times10^{11}\,h^{-1}{\rm M}_\odot$ (bottom) in the $90\,h^{-1}{\rm Mpc}$ box.
}
\end{table}

\subsection{The influence of mass loss on formation time}
\label{sec:massloss}

While typical analytic models of structure formation predict a monotonous
growth of halo mass over time, haloes do undergo interactions with other
haloes which can lead to tidal mass-loss.
The influence of such a mass-loss event on the measured formation time of a 
halo can be estimated as follows. 
Several studies of structure formation in a bottom-up scenario
\citep[e.g.][]{vandenBosch2002,Wechsler2002} have suggested that the bulk of haloes 
assembles the mass exponentially with redshift:
\begin{equation}
\label{eq:mass_assem}
M(z) = M(0)\,\exp(-\alpha z)
\end{equation}
where $\alpha$ is the mass-assembly rate. In this case, 
$z_{\rm form}=\alpha^{-1}\log 2$. Let us assume that the smooth mass accretion
history is interrupted at redshift $z_{\rm loss}$ by a single mass loss of
$M_{\rm loss}\equiv x\,M(z_{\rm loss})$. For simplicity, the growth rate
$\alpha$ is assumed to be identical before and after $z_{\rm loss}$ (it is
trivial to extend our results to the most general case). A short calculation
shows that mass loss leads to an increase in measured formation redshift given by
\begin{equation}
z_{\rm form} = \frac{1}{\alpha}\,\ln\,\frac{2}{1-x}.
\label{eq:change_zform}
\end{equation}

%%%%%% SECTION: TIDAL EFFECTS %%%%%%%%%%%%%%%%%%%%%%%%%%%%%%%%%%%%%%%%%%%%%%%%%%%
\section{The Effect of Tidal Forces on the Assembly of Halo Mass}
In this section we will demonstrate that the proximity to a massive halo
provides a special environment for the formation and evolution
of lower-mass haloes due to tidally sheared flows influencing their assembly.

\subsection{Halo evolution in the proximity to a massive halo}
\label{sec:proximity}
\begin{figure*}
\begin{center}
	\includegraphics[width=0.45\textwidth]{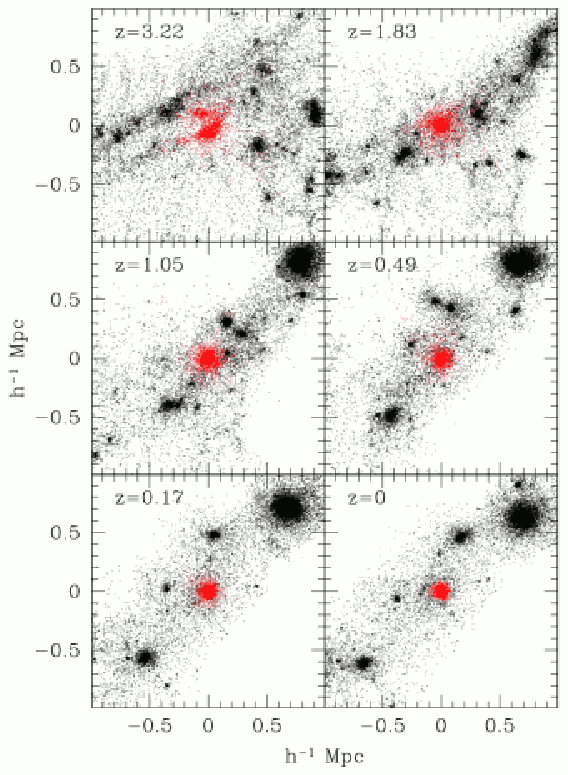}
	\hspace{0.03\textwidth}
	\includegraphics[width=0.45\textwidth]{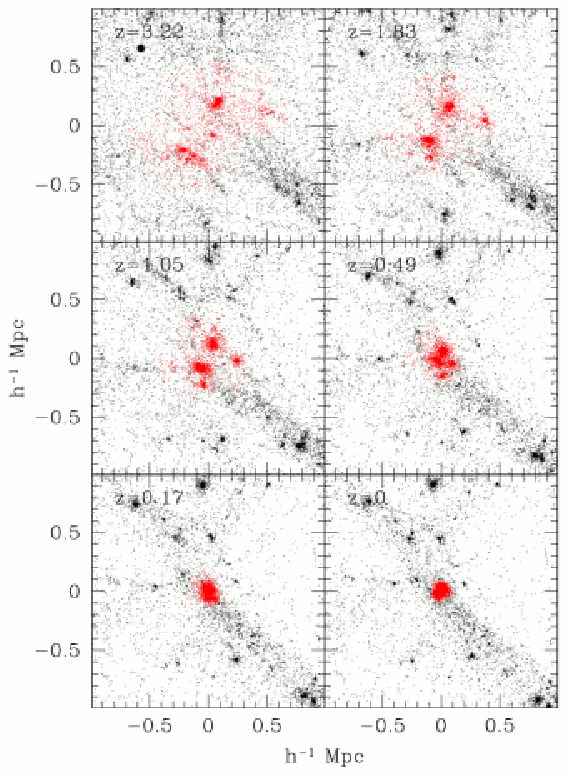}
\end{center}
\caption{\label{fig:HaloPatch} Evolution of the Lagrangian patch and environment
of two haloes with mass $1.4\times10^{11}\,h^{-1}{\rm M}_\odot$ at $z=0$ and
formation redshift $z_{\rm form}=3.1$ ({\it left}) and $z_{\rm form}=0.42$ ({\it
right}). Particles that will constitute the halo at $z=0$ (the Lagrangian patch)
are shown in red while particles that are contained in a $3\,h^{-1}{\rm Mpc}$
cube centred on the patch (the Lagrangian environment) are shown in black. The
centre of the coordinate system moves with the centre of mass of the particles
that constitute the halo at $z=0$.  }
\end{figure*}

For a large number of haloes in our simulation, we investigated the evolution
of their Lagrangian patches and of their Lagrangian environments.
At all epochs, the Lagrangian patch of a halo is defined by the set of 
particles that are identified as part of the halo at $z=0$. Similarly,
we define the Lagrangian environment as the set of particles that in the
initial conditions of the simulations were lying within a cube of 
$3  h^{-1}{\rm Mpc}$ centred on the corresponding Lagrangian patch.
We observe significant differences between old and young haloes of a given mass.

The older haloes are preferentially found in pronounced large-scale structure
already at high redshifts, while this is not true for the younger haloes. In
Figure \ref{fig:HaloPatch} we show the evolution of the particles making up two
different haloes at $z=0$ together with their Lagrangian environment. One of the
haloes has a very high formation redshift of $z_{\rm form}=3.1$ while the other
halo has a low formation redshift of $z_{\rm form}=0.42$. It is evident from the
evolution of the environment of the old halo that, once the host filament
formed, its neighbouring haloes show an expanding, ``Hubble-like'', motion along
the filament, which is not present for the younger halo. In the rest-frame of
the halo, this expansion along the filament appears as a recession of the
neighbouring haloes in that direction. Remarkably, in both cases, the last major
merger occurs perpendicular to the host large-scale structure. Thus, the
gravitational shear accelerates the collapse perpendicular to the filament while
it slows down the collapse along the filament
\citep[cf.][]{Porciani02,Porciani02b}. Tidal forces are expected to increase
substantially in the vicinity of very massive haloes leading to a strong
correlation between the tidal field and the density field on larger scales. This
leads to the correlation of halo properties that are influenced by the external
tidal field with the density field.

\subsection{Quantifying sheared flows}
\label{sec:tidalshearflows}
The dynamics of dark-matter flows (and thus the accretion onto dark-matter
haloes) is regulated by the gradient of the gravitational potential
$\nabla \phi$, its Laplacian $\nabla^2\phi$ and the tidal tensor
\begin{equation}
T_{ij}=\left[\frac{\partial^2}{\partial_i \partial_j}
-\frac{1}{3}\,\delta_{ij}\,\nabla^2 \right]\phi\;,
\end{equation}
where $\delta_{ij}$ denotes the Kronecker symbol. The presence of a massive halo
that locally dominates the gravitational potential generates a strong tidal
field in its neighbourhood. Over time, the tidal field produces a tidally
induced shear in the velocity field of the surrounding particles. The flow is
stretched in the direction of the massive halo and compressed in the plane
perpendicular to it. Since haloes preferentially reside in filaments or
sheets \citep[see e.g.][]{Hahn06}, tidal forces tend to accumulate coherently.
This should impact the mass accretion history of smaller haloes
neighbouring the massive one. To probe the influence of tidal forces, we
quantify the velocity field surrounding small-mass haloes by expanding the local
(peculiar) flow with respect to the bulk motion of the halo up to linear order
in the distance $x$ from the halo centre
\begin{equation}
\label{eq:vlinvelapprox}
v_i=\frac{\partial v_i}{\partial x_j} x_j.
\end{equation}
We next perform a least squares fit of the nine components of the
velocity gradient
tensor within one and four virial radii around each halo 
\citep[cf.][]{Porciani02,Porciani02b}. It is then
possible to write the velocity gradient tensor as the sum of a symmetric
and an antisymmetric part
\begin{eqnarray}
\frac{\partial v_i}{\partial x_j} &=& \frac{1}{2}\left(\frac{\partial
v_i}{\partial x_j}
+\frac{\partial v_j}{\partial
x_i}\right)+\frac{1}{2}\left(\frac{\partial v_i}{\partial x_j}
-\frac{\partial v_j}{\partial x_i}\right)\\
&\equiv&D_{ij}+\Omega_{ij},
\label{eq:definition_shearrot}
\end{eqnarray}
where $D_{ij}$ corresponds to the rate-of-strain tensor,
and $\Omega_{ij}$ describes rigid rotation due to nonzero vorticity
$\omega_i\equiv(\boldsymbol{\nabla}\times\mathbf{v})_i=-\epsilon_{ijk}\Omega_{jk}$,
where $\epsilon_{ijk}$ is the Levi-Civit\`a symbol.
The rate-of-strain tensor can be further decomposed as $D_{ij}\equiv
\theta \,\delta_{ij}+\Sigma_{ij}$ where $\theta$ denotes the velocity
divergence
(or expansion scalar) and $\Sigma_{ij}$ is the traceless shear tensor with eigenvalues 
$\mu_1\leq\mu_2\leq\mu_3$.
For growing-mode fluctuations in the linear regime, $T_{ij}\propto \Sigma_{ij}$
and therefore the eigenvectors of these matrices are aligned. However, the
proportionality of the shear and tide is broken by higher-order terms. In
general, the eigenvectors of the tensors remain parallel but the eigenvalues
evolve differently with time \citep[e.g.][]{Bertschinger1994}\footnote{This
holds in the Lagrangian frame before shell-crossing \citep[see also][]{Barnes1989}.}. 
In fact, while tides act as sources for
the velocity shear, the latter is also influenced by the velocity gradient,
vorticity and the shear itself. In what follows, we use the velocity shear as a
measure of the integrated effect of the gravitational tides after 
investigating the connection between the two in Section \ref{sec:rhill}.

Both random motion and nonlinear ordered motion will be superimposed onto the
ordered linear deformation, as defined in eq. (\ref{eq:vlinvelapprox}), 
so that we can define a velocity dispersion along the
three main axes of the rate-of-strain tensor quantifying this residual by
\begin{equation}
\sigma_i^2 \equiv \left\langle\left(
(\mathbf{v}-\lambda_{i}\mathbf{x}-\boldsymbol{\omega}\times\mathbf{x})\cdot\mathbf{w}_i
\right)^2\right\rangle,
\end{equation}
where $\lambda_1\leq\lambda_2\leq\lambda_3$ are the eigenvalues and
$\mathbf{w}_i$ are the corresponding eigenvectors of the rate-of-strain tensor
$D_{ij}$. Furthermore, we define a simple relative velocity $\mathbf{v}_{\rm
rel}$ between a halo's centre of mass velocity and the mean velocity of the
region between one and four virial radii. If the velocity field is completely
linear then both $\left|\mathbf{v}_{\rm rel}\right|=0$ and
$\left|\boldsymbol{\sigma}\right|=0$. We observe that, apart from very little
scatter,  $\left|\boldsymbol{\sigma}\right|$ is completely dominated by
$\left|\mathbf{v}_{\rm rel}\right|$. Nonzero $|\boldsymbol{\sigma}|$ and
$|\mathbf{v}_{\rm rel}|$ can arise when the velocity grows non-linearly with
distance from the halo, due to the presence of other haloes within $1-4$ virial
radii or in special circumstances when the halo has indeed a relative velocity
with respect to its surrounding flow. We will return to this last aspect in
Section \ref{sec:sim_stripping}. Furthermore, noise in the velocity field due to
finite numerical resolution will also increase $|\boldsymbol{\sigma}|$ by
causing spurious deviations from a flow with constant strain and thus impacting the
goodness of the fit (see also Section \ref{sec:tidesinfluence}).

%%% new material

\subsection{The tidal origin of sheared flows}
\label{sec:rhill}

\begin{table}
\begin{center}
\begin{tabular}{lccc}
\hline
\multicolumn{1}{c}{$\rho_s$}   & $z=0$ 	& $z=0.5$ 	& $z=1$ \\
\hline
$\left(\lambda_3,M/d^3\right)$ & 0.62 & 0.64 & 0.61 \\
$\left(\mu_3,M/d^3\right)$     & 0.75 & 0.64 & 0.57 \\
\hline
\end{tabular}
\end{center}
\caption{\label{tab:corr_shear}Spearman rank correlation coefficients $\rho_s$
between the largest tidal field eigenvalue due to the tidally dominant neighbour and
the strain eigenvalues $\lambda_3$ and shear eigenvalues $\mu_3$, respectively.
The coefficients are given for haloes identified at $z=0$ with masses between 2
and $4\times10^{10}\,h^{-1}{\rm M}_\odot$.}
\end{table}

\begin{figure*}
\begin{center}
	\includegraphics[width=0.45\textwidth]{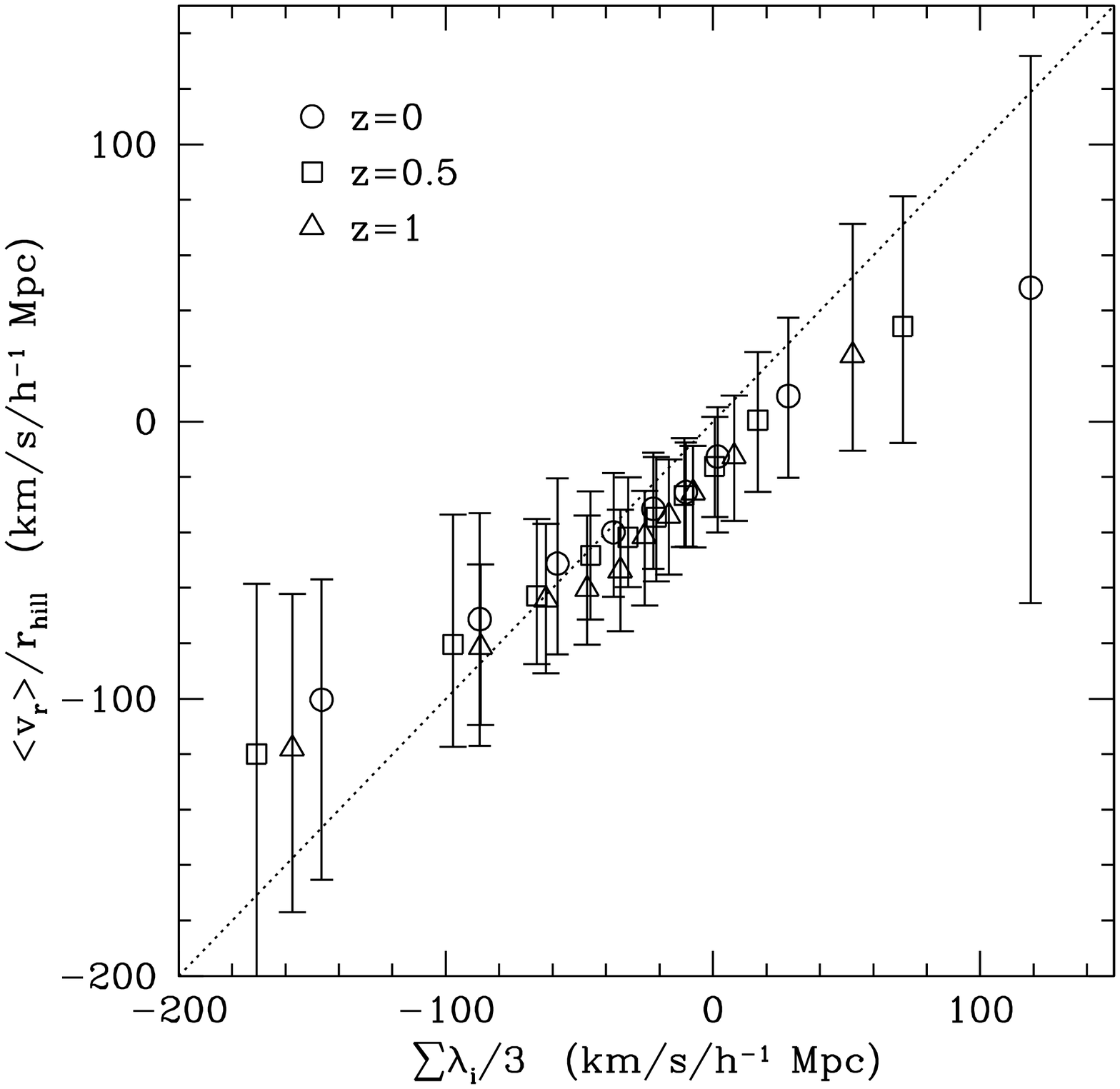}
	\hspace{0.03\textwidth}
	\includegraphics[width=0.45\textwidth]{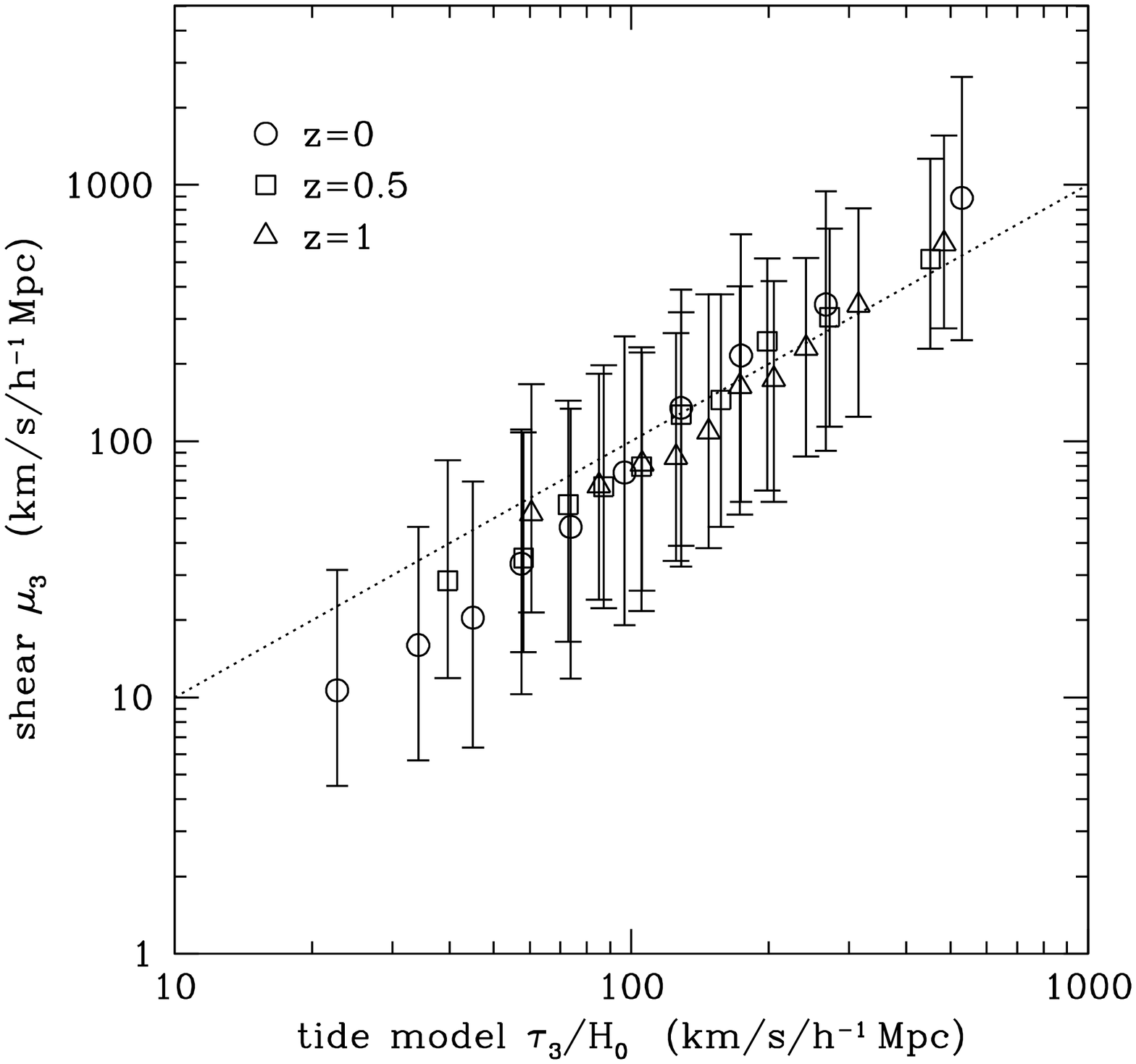}
\end{center}
\caption{\label{fig:tide_shear}
Comparisons between the flux onto haloes from particle velocities and in 
the CSA model, and between the tides due to the tidally dominant neighbour
halo and measured shear in the CSA. (Left panel:) The flow through the Hill
sphere $\sum_i\lambda_i/3$ from the linear fit to the flow against median
$\left<v_r\right>/r_{\rm Hill}$ measured directly from the simulation. 
Results are shown for haloes with $r_{\rm Hill}\leq 4r_{\rm vir}$, the
region used for determining the $\lambda_i$. (Right panel:) Median eigenvalue 
of the shear tensor $\mu_3$ measured from the flow as a function of the tidal 
force due to the tidally dominant neighbour halo $\tau_3=2GM/R^3$. Errorbars 
show the 16th and 84th percentiles in each bin. Bins are chosen to include 
equal numbers of datapoints. The dotted lines indicate $x=y$. Data are given 
for the progenitors of haloes with masses between $2$ and 
$4\times10^{10}\,h^{-1}{\rm M}_\odot$.
}
\end{figure*}

In this section we will provide evidence that the tidal field due to a
neighbouring massive halo induces shear in the dark matter flow surrounding
smaller haloes in its vicinity. To this end, we will compare the constant strain
approximation (CSA in what follows) introduced in the previous section,
i.e. that the flow is described by $D_{ij}$ in eq. (\ref{eq:definition_shearrot}), 
to a restricted three-body approximation for the tidal field. This will allow us 
to use the shear measured from the flow to quantify the importance of tides on the 
mass assembly of haloes in the rest of the paper.

Consider the restricted three-body problem, consisting of a larger body of mass
$M$, a smaller body of mass $m\ll M$ and a third of negligible mass. The third
body can have stable circular orbits around the smaller mass $m$ only within the
Hill radius
\begin{equation}
r_{\rm Hill} \simeq d\left(\frac{m}{3M}\right)^{1/3},
\label{eq:hillradius}
\end{equation}
where $d$ is the distance between $m$ and $M$ 
\citep[for a detailed derivation we refer the reader to][]{Murray2000}. Since in
cosmological context we do not have the idealized situation of the restricted
three-body problem, we define the Hill radius of each halo as the minimal Hill
radius due to any of its neighbours\footnote{This will possibly overestimate
the Hill radius. The restricted three-body problem applies to the orbit of a 
test particle of zero mass around the mass $m$. While in the case of smooth 
accretion the third body can indeed be thought of as having negligible mass, 
in the case of mergers, the restricted three-body problem is of course of
limited applicability.

Furthermore, the presence of finite masses other than $m$ and $M$ makes it 
unfortunately impossible to use radial acceleration vectors to find the region 
of gravitational influence of the second body $m$. Any self-bound mass (e.g. 
subhaloes or other tiny haloes outside the virial radius) will produce a dip in the 
potential. In order to use acceleration vectors it would be necessary to remove 
all structure smaller than the halo itself, which is practically impossible. 
In fact, for the same reasons we consider the shear tensor rather than the 
tidal tensor since the latter is affected by the same problems. Shear is the 
integrated result of tidal forces and thus probes directly their effect on the 
flow at the relevant scales.}.
Now, given that the eigenvalues of the
tidal tensor $T_{ij}$ for a point mass $M$ at distance $d$ are given by
\begin{equation}
\left\{\tau_i\right\} \equiv \left\{ -\frac{GM}{d^3},\,-\frac{GM}{d^3},\,\frac{2GM}{d^3}\right\},
\end{equation}
this choice is equivalent to selecting that neighbour which has maximal tidal 
influence $GM/d^3$  -- the tidally dominant neighbour halo.
The eigenvector $\mathbf{t}_3$ associated to the third eigenvalue $\tau_3$ 
points radially towards the mass $M$, while $\mathbf{t}_1$ and $\mathbf{t}_2$
span the perpendicular plane but are otherwise undetermined due to the symmetry
of the problem.

By investigating the flux through the Hill sphere (with volume $V_\textrm{Hill}$ 
and surface $\partial V_\textrm{Hill}$), it is then possible
to connect the sheared motion of the flow around a halo, measured using
the CSA defined in Section \ref{sec:tidalshearflows}, to the 
tidal forces due to the neighbouring halo. This flux is given by
\begin{eqnarray}
\Phi & \equiv & \int_{\partial V_{\rm Hill}} \mathbf{v}\cdot{\rm d}\mathbf{A} \\
& = &\int_{V_{\rm Hill}} \boldsymbol{\nabla}\cdot\mathbf{v}\, {\rm d}V
 \simeq \frac{4\pi}{3}r_{\rm Hill}^3\,(\lambda_1+\lambda_2+\lambda_3),
 \label{eq:fluxmodel1}
\end{eqnarray}
where the first equality is just an application of the divergence theorem and 
the second follows after applying the CSA to the flow where
the $\lambda_i$ are assumed to be constant within $r_{\rm Hill}$.

\begin{figure}
\begin{center}
  \includegraphics[width=0.45\textwidth]{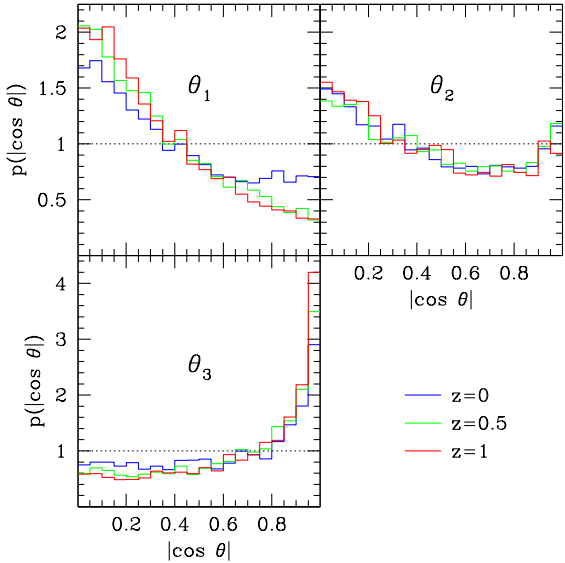}
\end{center}
\caption{\label{fig:tide_shear_angle} The probability distributions of angles $|\cos \theta_i|$ between
the three rate-of-strain eigenvectors $\mathbf{w}_i$ and the direction to the tidally
dominant neighbour halo for the progenitors of haloes with $z=0$ masses between $2$ and
$4\times10^{10}\,h^{-1}{\rm M}_\odot$ for three different redshifts: $z=0$ (blue),
$z=0.5$ (green) and $z=1$ (red). The black dotted line indicates a random distribution
of angles.}
\end{figure}

In order to verify that the shear measured in the flow indeed satisfies the
CSA, we can rewrite eq. (\ref{eq:fluxmodel1}) as
\begin{equation}
\frac{\left<v_r\right>}{r_{\rm Hill}}\simeq\frac{1}{3}\left(\lambda_1+\lambda_2+\lambda_3\right),
\label{eq:fluxtest}
\end{equation}
where $\left<v_r\right>$ is the mean radial velocity at $r_{\rm Hill}$. Written
this way, the left hand side is independent of the assumption of constant strain. 
We can thus verify the validity of equation (\ref{eq:fluxtest}) directly in the
$N$-body simulations in order to assert that the CSA is
justified. In Figure \ref{fig:tide_shear} we show $\left<v_r\right>/r_{\rm
Hill}$ as a function of the mean of the strain eigenvalues for haloes with
$r_{\rm Hill}<4r_{\rm vir}$, the range used when fitting the CSA flow model.
The mean radial velocity was measured using the closest 100 particles outside
the Hill sphere. Results are shown for haloes that have final masses at $z=0$
between $2-4\times10^{10}\,h^{-1}{\rm M}_\odot$. The rank correlation is
$\rho_s(\left<v_r\right>/r_{\rm Hill},\sum_i\lambda_i/3)=0.74$. This very strong 
correlation indicates that the constant strain model serves well to predict the 
flux into the Hill sphere.

In addition, we can check directly whether $\lambda_3$ is probing the tidal
influence due to a neighbouring halo in which case we expect that
$\lambda_3\propto\tau_3$. Since $\tau_3\propto M/d^3$ is a positive definite
quantity, we show the shear eigenvalue $\mu_3$, which has the same property, as
a function of $M/d^3$ in the right panel of Figure \ref{fig:tide_shear}. Results
are given at three redshifts for the progenitors of haloes with masses between
$2$ and $4\times10^{10}\,h^{-1}{\rm M}_\odot$. We find a tight connection
between the two quantities which is also reflected in the high rank correlations
given in Table \ref{tab:corr_shear}. 

Finally, we can also verify whether the tidal field eigenvector $\mathbf{t}_3$,
which is simply the normalized vector pointing radially towards the tidally dominant
halo, is parallel to the third eigenvector $\mathbf{w}_3$ of the strain tensor
and perpendicular to the first and the second, as would be expected if tides are
inducing shear in the flow. Figure \ref{fig:tide_shear_angle} 
shows the distribution of the cosine of the angles between these vectors, i.e.
\begin{equation}
\left|\cos \theta_i\right| = \left|\frac{\mathbf{w}_i \cdot \mathbf{t}_3}{\left\| \mathbf{w}_i \right\| }\right|
\end{equation}
at three redshifts in the simulation. If the strain on the fluid is indeed caused by the 
tidal field due to the neighbouring halo, then an alignment angle $|\cos \theta_3|\simeq 1$ 
is expected which is consistent with the results shown in Figure \ref{fig:tide_shear_angle}.
It is interesting to see that the degree of alignment -- reflected in the width of the distributions
-- is slightly decreasing with decreasing redshift.

The findings presented in this section -- in particular (1) that
the rate-of-strain eigenvalue $\lambda_3$ is very strongly correlated with 
the tidal field eigenvalue due to the tidally dominant neighbour halo and 
(2) that the associated rate-of-strain eigenvector points into the direction 
of this halo -- lead us to conclude that $\lambda_3$ can be used as a proxy 
to quantify the integrated effect of tides on the flow surrounding haloes 
in our subsequent analysis.

\subsection{The tidal influence on mass assembly}

\begin{figure*}
\begin{center}
	\includegraphics[width=0.3\textwidth]{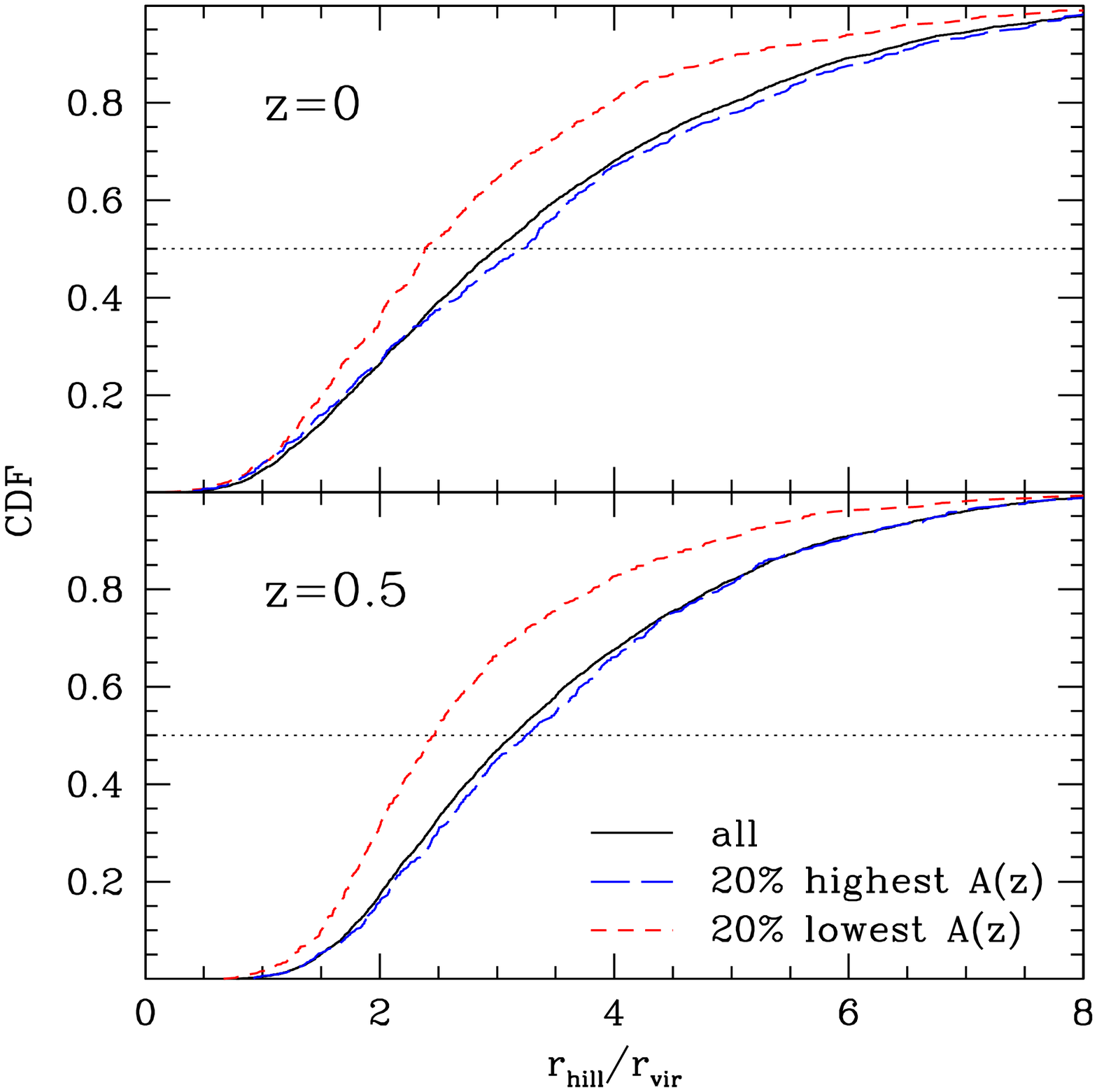}
  \hspace{5mm}
  \includegraphics[width=0.3\textwidth]{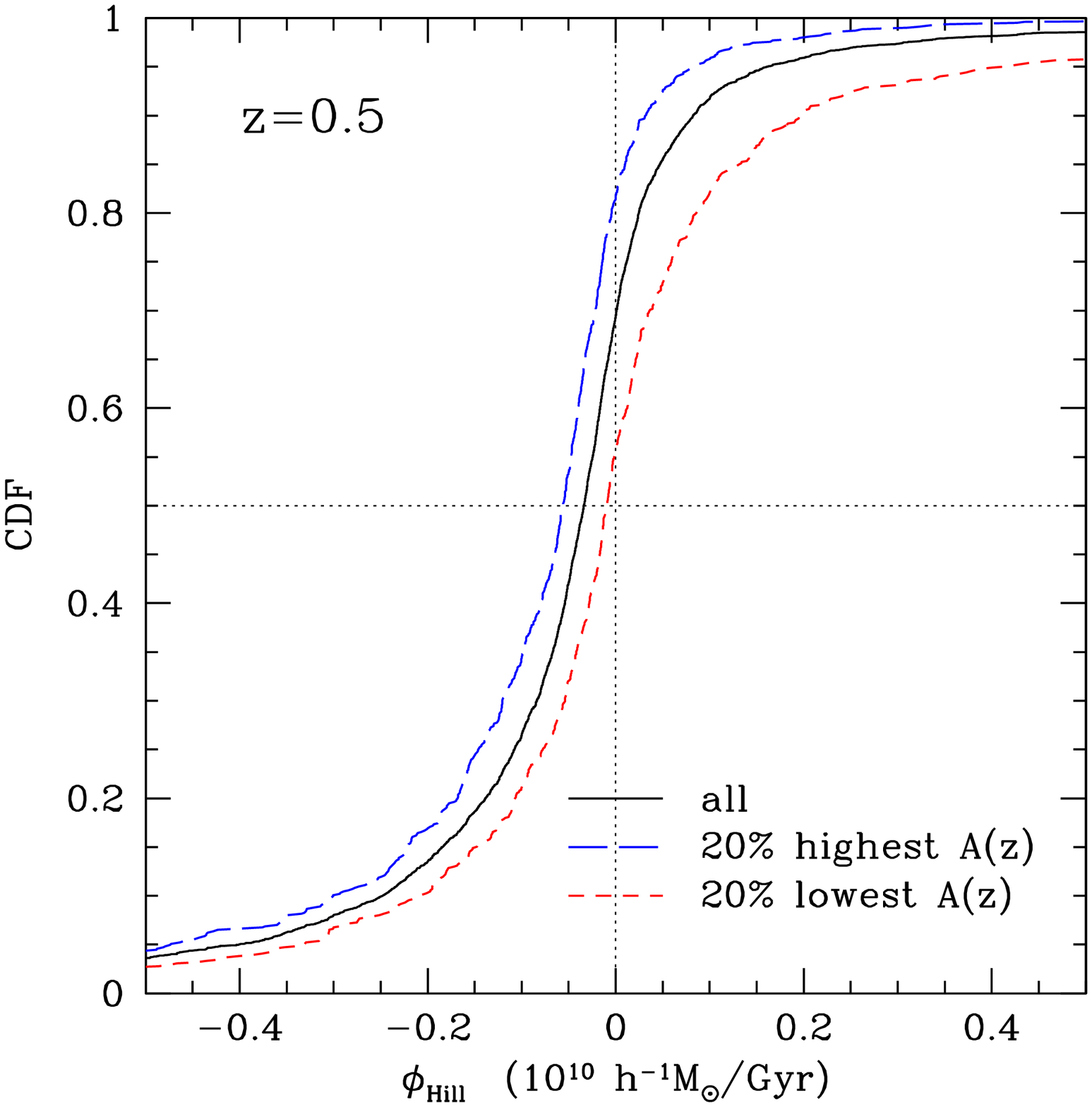}
  \hspace{5mm}
  \includegraphics[width=0.3\textwidth]{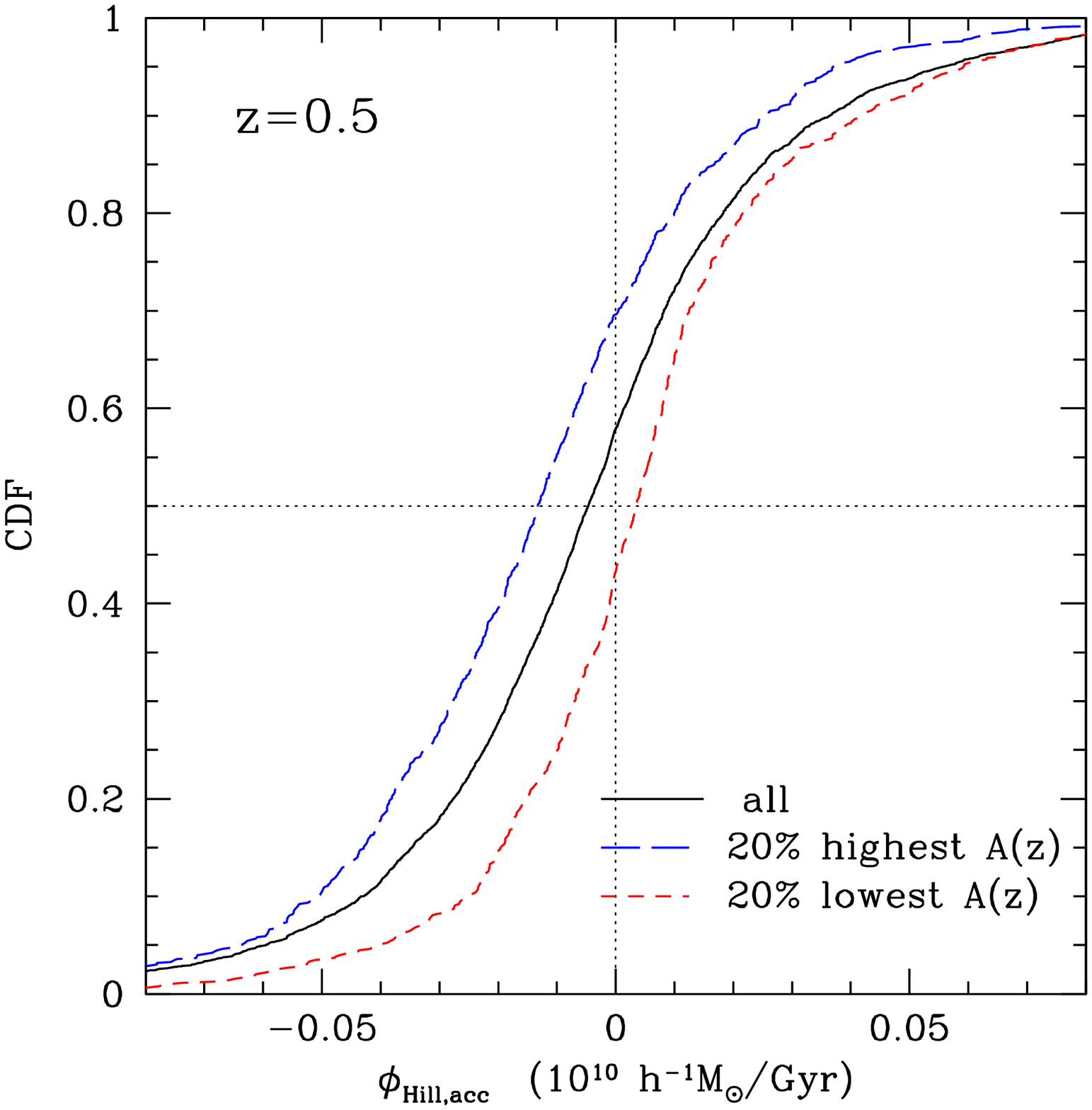}
\end{center}
\caption{\label{fig:assemb_tides}({\it Left}) Cumulative distribution 
functions (CDF) of Hill radii in units of virial radii, and ({\it Middle}) 
CDFs of the mass flux through the Hill sphere, ({\it Right}) same as
middle panel but only for low-velocity particles with $v<v_{\rm vir}$. All data are 
shown for the progenitors at $z=0.5$ of haloes 
with final masses between $2$ and $4\times10^{10}\,h^{-1}{\rm M}_\odot$ (and also at 
$z=0$ for the left panel). The CDFs are shown in black for the sample of 
all these haloes and in red (blue) for those haloes with the 20 per cent lowest 
(highest) assembly rate $A(z)$. Stripped haloes are excluded. }
\end{figure*}

The assembly of haloes is regulated by the amount
of accretable material in their vicinity. The sphere of gravitational influence of 
a halo is bounded by the Hill sphere, only particles within it are potentially
accretable. We will in this section quantify differences in the supply of accretable
material and its relation to mass growth and tides.

In Figure \ref{fig:assemb_tides}, 
left panel, we show the cumulative distribution of Hill radii (in units of virial 
radii) for the $z=0.5$ progenitors of haloes with masses between $2$ and 
$4\times10^{10}\,h^{-1}{\rm M}_\odot$ and for these haloes at z=0. The data are 
shown for all haloes and for those with the 20 per cent highest and lowest current 
assembly rates $A(z)$. We find that at both redshifts the estimated Hill radii of 
the least accreting haloes are significantly smaller than those of the most 
accreting haloes, which are comparable to those of all haloes. This implies that 
haloes with the lowest assembly rates experience stronger tidal fields than all 
other haloes. 

The mass assembly rate $A(z)$, which can also be thought of as the 
mass flux through the halo surface
\begin{equation}
\frac{{\rm d}M_h}{{\rm d}t} = \int_{\partial V_{\rm vir}}\rho\,\mathbf{v} \cdot {\rm d}\mathbf{S},
\end{equation}
is expected to be correlated through the continuity equation to the supply rate of 
accretable material. The latter is the mass flux through the Hill sphere
\begin{equation}
\phi_{\rm Hill} \equiv \int_{\partial V_{\rm Hill}}\rho\, \mathbf{v} \cdot {\rm d}\mathbf{S}.
\end{equation}
In both definitions $\rho$ is the matter density, $M_h$ is the halo mass.
This correlation is however likely to be weakened by several factors. A particle 
does not enter the Hill sphere 
and the virial radius simultaneously, especially when the Hill radius is large,
leading to a time delayed correlation with the delay depending on the size of $r_{\rm Hill}$.
Furthermore, the Hill radius shrinks linearly with the distance to a more massive
halo (see eq. \ref{eq:hillradius}). In addition, the tidal field will perturb 
particle trajectories also inside the Hill sphere. Thus, entering the Hill sphere is a 
necessary but not sufficient condition for a particle to be accreted in the future. 
Nevertheless, we can compare the flux through the Hill sphere for haloes with high 
and low assembly rates to probe the supply of
accretable material and its relation to the assembly behaviour of haloes. 
In Figure \ref{fig:assemb_tides}, middle panel, we show 
cumulative distributions of the flux through the Hill sphere for haloes with highest 
and lowest current assembly rates $A(z)$. The flux was computed using the closest 100 
particles outside the Hill sphere from 
\begin{equation}
\phi_{\rm Hill}\simeq4\pi\,R_{\rm Hill}^2\,\left<v_r\,\rho\right>,
\end{equation}
where $v_r$ is the radial velocity and $\left<\cdot\right>$ denotes the average over
the shell occupied by the 100 particles. We find that the Hill sphere flux of the least
accreting haloes is shifted towards more positive values relative to
the most accreting sample, i.e. the flux into the Hill sphere is
reduced for the least accreting haloes. In particular, these haloes have a 
median flux very close to zero.

Particles that enter the Hill sphere might not be accretable if they
have too high velocities to be captured by the halo \citep[see also][]{wang06}. 
In the plane perpendicular to the eigenvector belonging to $\tau_3$ the tidal 
field compresses the flow and thus accelerates particles onto the halo. Furthermore, 
high velocities could also be due to particles orbiting the bigger halo which has 
a larger circular velocity than the smaller halo. To
quantify the abundance of these particles that are potentially too fast for the
halo to be captured, we determine the fraction of particles entering the Hill
sphere that have a velocity smaller than the circular velocity. This is a
conservative requirement since the halo should be able to capture particles
up to its escape velocity (which is twice its circular velocity).
We thus recompute the flux through the Hill sphere using only these accretable
particles. In Figure \ref{fig:assemb_tides}, right panel, the cumulative 
distributions of accretable fluxes are shown for the same halo samples as before. 
We find that the difference between the distributions increases slightly, the median 
flux through the Hill sphere is still very close to zero (if not slightly positive) 
for the least accreting haloes while it is negative for the most accreting haloes.

These results suggest that accretion is not primarily limited by high velocities 
since the difference between the flux distributions does not change substantially
when considering all particles as opposed to only slow particles. Hence,
non-accreting haloes do not even draw on a supply of slow particles suggesting
that they are no longer convergence points in the local flow of dark matter
particles\footnote{Note that the relatively low flux through the Hill sphere for the 
most-accreting haloes appears to be in conflict with their assembly rates and 
thus the growth they should experience. This incongruity is however easily resolved
by considering the relative sizes of the Hill radii (see Figure \ref{fig:assemb_tides}). 
The flux is constrained to the assembly rate at the virial radius and to 
$\phi_\textrm{ta}\simeq-32\pi\bar{\rho}H(z)r_{\rm vir}^3\simeq-9\times10^{11}
\left(\frac{r_\textrm{vir}}{h^{-1}\textrm{Mpc}}\right)^3h^{-1}\textrm{M}_\odot/\textrm{Gyr}$
at the turnaround radius (assuming that the density at the turnaround radius is $\bar{\rho}$,
that $r_\textrm{ta}=2r_\textrm{vir}$ and velocity is $-H(z)r_\textrm{ta}$).
For the haloes we consider, $r_\textrm{vir}\simeq0.09\,h^{-1}\textrm{Mpc}$ and 
so $\phi_\textrm{ta}\simeq -7\times 10^{8}\,h^{-1}\textrm{M}_\odot/\textrm{Gyr}$. Thus, if all the 
material that the halo will accrete 
until $z=0$ is already within its Hill sphere, then the Hill radius will be on 
the order of or larger than the turnaround radius. Given that the overdensity 
within the turnaround radius in the spherical collapse model is found to be 4.56,
we can simply compare the overdensity within the estimated Hill radius with this
number to compare their relative sizes. We find that the median overdensities within 
the Hill sphere for the haloes with the 20 per cent highest and lowest assembly rate 
$A(z)$ at $z=0.5$ are 5.54 and 12.72, respectively. Thus, for the high-$A(z)$
sample, the relatively low Hill flux is perfectly compatible with their large 
assembly rates as this is the flux determined close to the turnaround radius. 
Since their estimated Hill radii are comparable to their turnaround radii, the growth 
of these haloes is simply unaffected by tides. Indeed, 94 per cent of the high-$A(z)$
and 99 per cent of the low-$A(z)$ sample have at least their final mass contained 
inside their estimated Hill radius at $z=0.5$.
Note also that in dark-energy cosmologies the value for the turnaround overdensity 
at low redshifts can be significantly higher than the fiducial Einstein-de~Sitter value 
\citep[see e.g.][]{Horellou2005}.}.

%%% end of new material

\subsection{The shear flow near early forming haloes}
\label{sec:shearflow}
\begin{figure}
\begin{center}
	\includegraphics[width=0.45\textwidth]{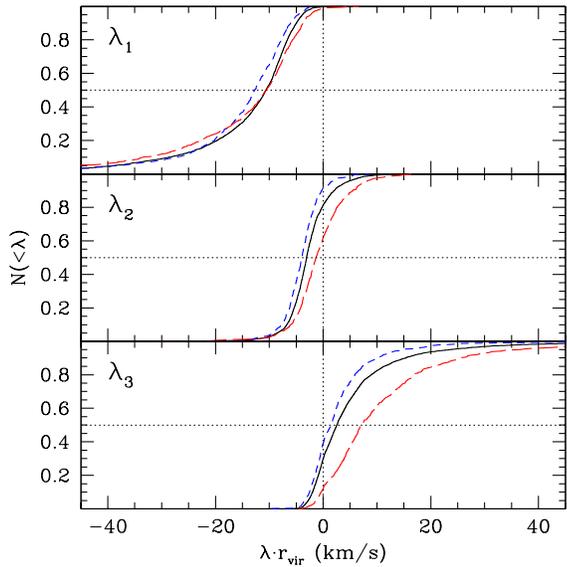}
\end{center}
\caption{\label{fig:ecdf_shear} Cumulative distribution functions of the
eigenvalues $\lambda_i$ of the deformation tensor $D_{ij}$. The data is shown for 
the $z=0.5$ progenitors of haloes with
masses between $2\times10^{10}$ and $4\times10^{10}\,h^{-1}{\rm M}_\odot$ at
$z=0$. Lines correspond to all these haloes (solid, black), those 20 per cent
with highest current assembly rate (short dashes, blue) and those with the 
lowest (long dashes, red). Stripped haloes are excluded.}
\end{figure}

We will now discuss the connection between a sheared velocity field surrounding 
low mass haloes and their mass assembly properties, thus providing further
evidence for an influence of external tides on mass assembly.

In Figure \ref{fig:ecdf_shear}, we show the cumulative distribution functions
of the eigenvalues $\lambda_i$ of the deformation tensor $D_{ij}$ for 
haloes with highest and lowest assembly rates $A$ at $z=0.5$, where the effect 
is strongest. We observe a systematic shift of $\lambda_2$ and $\lambda_3$
towards more positive values for those haloes that grow slower. This 
implies that the flow around these haloes is more strongly sheared. 
As discussed in Section \ref{sec:rhill}, sheared motion in the flow provides 
clear evidence for the presence of an external tidal field. In particular,
for the low-$A$ haloes, over 80 per cent have $\lambda_3>0$, i.e. the
velocity component along the eigenvector associated with $\lambda_3$
is positive, particles are moving away from the halo in that direction.

\begin{figure}
\begin{center}
	\includegraphics[width=0.45\textwidth]{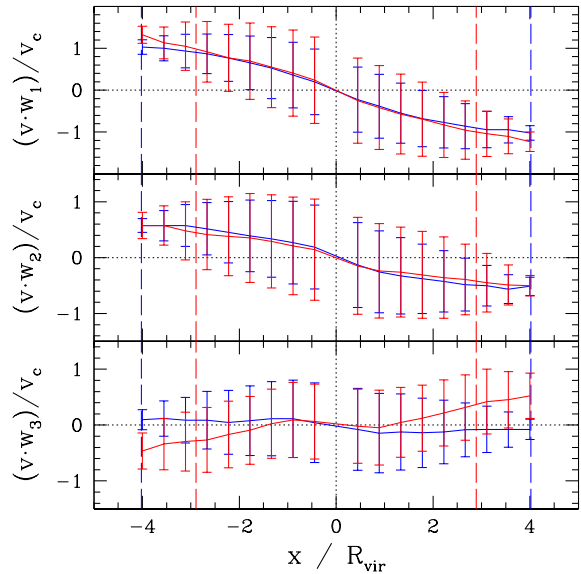}	
\end{center}
\caption{\label{fig:velfield}Median velocity field along the three main axes
of the rate-of-strain tensor $\mathbf{w}_1$ (top), $\mathbf{w}_2$ (middle)
and $\mathbf{w}_3$ (bottom) in units of the halo circular velocity $v_c$. 
The data is shown for the $z=1$ progenitors
of haloes with masses between $2\times10^{10}$ and 
$4\times10^{10}\,h^{-1}{\rm M}_\odot$ at $z=0$.
The sample of haloes is split into those 20 per cent with largest (blue) 
and smallest (red) increase in mass. Stripped haloes are excluded. Errorbars 
correspond to the medians (over all haloes) of the 16th and 84th percentiles 
(computed for each halo). 
The vertical dashed lines indicate the median Hill radii for the
two samples of haloes.}
\end{figure}

In Figure \ref{fig:velfield} we show the median velocity of the environment as a
function of distance from the halo centre projected along the three main axes of
the rate-of-strain tensor (which coincide with the main axes of the shear
tensor). Results are shown for haloes growing most and least in mass between
redshift one and zero for the progenitors of haloes of fixed mass at $z=0$. The
velocity is given in units of the circular velocity of each halo
$v_c\equiv\sqrt{GM/R_{\rm vir}}$. While the flow fields are comparable along
$\mathbf{w}_1$ and $\mathbf{w}_2$ apart from slightly larger scatter for the
less growing haloes, these haloes show a qualitatively different flow along the
eigenvector $\mathbf{w}_3$ corresponding to the largest eigenvalue $\lambda_3$.
For the least growing haloes, along this direction, matter is receding from the
halo centre leading to a Hubble-like flow that counteracts infall onto the halo.
The data are shown at $z=1$ where the difference in the flow field is strongest
when splitting the halo sample by growth $G(z)$. The scatter shown in Figure 
\ref{fig:velfield}
indicates that there is a non-negligible fraction of particles surrounding the
halo that have velocities on the order of the halo's circular velocity, as
discussed and quantified in the previous section. Note that outflow along
one direction is equivalent to a non-convergent velocity field. Such a non-convergent
flow is in excellent agreement with other detections of matter outflows around haloes
below the non-linear mass \citep[e.g][]{Prada2006}.

%%%% SECTION: TIDAL ORIGIN OF ASSEMBLY BIAS  %%%%%%%%%%%%%%%%%%%%%%%%%%%%%
\section{The role of tidal effects in the assembly bias}
\label{sec:TidalAssembly} The hierarchical structure-formation picture based on
the Extended Press-Schechter model predicts that the mass of the main progenitor
of a halo is determined by the sequence of ``density peaks'' on increasingly
larger scales that exceed the threshold for collapse. In this picture, formation
time depends only on the halo mass without any influence of the environment.\\
We will argue in this section that in $N$-body simulations the 
assembly behaviour of haloes is modulated by tidal effects ranging from the 
suppression of halo growth to tidal mass loss due to an encounter with a massive 
halo in the most extreme cases.

\begin{table}
\begin{center}
\begin{tabular}{lccc}
\multicolumn{1}{c}{$\rho_s$}   & $z=0$ 	& $z=0.5$ 	& $z=1$ \\
\hline
$\left(z_{\rm form},\lambda_3(z)\right)$ & \textbf{0.11} & \textbf{0.22} & \textbf{0.30} \\
$\left(z_{\rm form},\delta_2(z)\right)$  & 0.15 & 0.17 & 0.19 \\
$\left(z_{\rm form},|\boldsymbol{\sigma}(z)|\right)$  & 0.15 & 0.18 & 0.22 \\
\hline
$\left(A(z),\lambda_3(z)\right)$            & \textbf{-0.13} & \textbf{-0.28} & \textbf{-0.23} \\
$\left(A(z),\delta_2(z)\right)$             & -0.18 & -0.18 & -0.15 \\
$\left(A(z),|\boldsymbol{\sigma}(z)|\right)$             & -0.17 & -0.20 & -0.14 \\
\hline
$\left(G(z),\lambda_3(z)\right)$ & \textbf{-} & \textbf{-0.29} & \textbf{-0.32} \\
$\left(G(z),\delta_2(z)\right)$  & - & -0.13 & -0.18 \\
$\left(G(z),|\boldsymbol{\sigma}(z)|\right)$  & - &  -0.19 &  -0.21 \\
\hline
$\left(\delta_2(z),\lambda_3(z)\right)$ & 0.60 & 0.55 & 0.47 \\
\hline
\hline
$M_{\rm med}(z)$ & $2.7$ & $2.2$ & $1.7$ \\
\hline
\end{tabular}
\end{center}
\caption{\label{tab:corr_lambda}Spearman rank correlation coefficients $\rho_s$
between measures of halo mass assembly and measures of environment over
redshift. Measures of environment are $\lambda_3$, the largest eigenvalues of
the rate-of-strain tensor, the overdensity $\delta_S$ in the shell within one
and four virial radii around each halo, $\delta_2$ the matter field smoothed
with a spherical top-hat filter of scale $2\,h^{-1}\,{\rm Mpc}$ and the
deviation from a linear flow $|\boldsymbol{\sigma}|$. Measures of mass assembly
are the formation redshift $z_{\rm form}$, the assembly rate $A(z)$ and the mass
ratio $G(z)$. The coefficients are given for haloes with masses between $2$ and
$4\times10^{10}\,h^{-1}{\rm M}_\odot$ at $z=0$. These haloes have a median
formation redshift $z_{\rm med}=1.71$. The median mass $M_{\rm med}(z)$ of the
halo sample at redshift $z$ is given in units of $10^{10} h^{-1}{\rm M}_\odot$.
Data is obtained from the $45\,h^{-1}{\rm Mpc}$ box. Tidally stripped haloes are
excluded. These results illustrate that: (1) mass assembly is modulated by
$\lambda_3$ and (2) the correlation of $\lambda_3$ with density grows over time.}
\end{table}

\begin{table}
\begin{center}
\begin{tabular}{lccc}
\hline
\multicolumn{1}{c}{$\rho_s$}   & $z=0$ 	& $z=0.5$ 	& $z=1$ \\
\hline
$\left(z_{\rm form},\lambda_3(z)\right)$ & \textbf{0.09} & \textbf{0.22} & \textbf{0.27} \\
$\left(z_{\rm form},\delta_2(z)\right)$  & 0.15 & 0.17 & 0.18 \\
$\left(z_{\rm form},|\boldsymbol{\sigma}(z)|\right)$  & 0.11 & 0.13  & 0.17 \\
\hline
$\left(A(z),\lambda_3(z)\right)$            & \textbf{-0.08 } & \textbf{-0.19 } & \textbf{-0.08 } \\
$\left(A(z),\delta_2(z)\right)$             & -0.13 & -0.14 & -0.07\\
$\left(A(z),|\boldsymbol{\sigma}(z)|\right)$& -0.12 & -0.13 & 0.0  \\
\hline
$\left(G(z),\lambda_3(z)\right)$ & \textbf{-} & \textbf{-0.24} & \textbf{-0.31} \\
$\left(G(z),\delta_2(z)\right)$  & - & -0.12 & -0.16\\
$\left(G(z),|\boldsymbol{\sigma}(z)|\right)$  & - &  -0.11 &  -0.15 \\
\hline
$\left(\delta_2(z),\lambda_3(z)\right)$ & 0.65 & 0.60 & 0.53 \\
\hline
\hline
$M_{\rm med}(z)$ & $27$ & $21$ & $16$\\
\hline
\end{tabular}
\end{center}
\caption{\label{tab:corr_lambda2}Same as Table \ref{tab:corr_lambda} but for 
haloes with masses between $2$ and $4\times10^{11}\,h^{-1}{\rm M}_\odot$ at 
$z=0$. These haloes have a median formation redshift $z_{\rm med}=1.44$.
Data is obtained from the $90\,h^{-1}{\rm Mpc}$ box. }
\end{table}

\subsection{The influence of tides on halo mass assembly}
\label{sec:tidesinfluence}
In order to statistically quantify the interplay between environment and halo mass assembly,
we present Spearman rank correlation coefficients between measures of
environment and measures of halo mass assembly in Tables \ref{tab:corr_lambda}
and \ref{tab:corr_lambda2} at three redshifts for the main progenitors of haloes
identified at $z=0$. In Table \ref{tab:corr_lambda}, data is shown for haloes in
the $45\,h^{-1}{\rm Mpc}$ simulation with masses between 2 and
$4\times10^{10}\,h^{-1}{\rm M}_\odot$ at $z=0$. Table \ref{tab:corr_lambda2}
shows the corresponding data for haloes with masses between 2 and
$4\times10^{11}\,h^{-1}{\rm M}_\odot$ at $z=0$ from the $90\,h^{-1}{\rm Mpc}$
simulation.

We find that for both mass ranges the correlation of formation redshift and the
eigenvalue $\lambda_3$ is strongest at redshifts close to the median formation
redshift of the sample. Furthermore, the correlation of assembly rate $A(z)$
with $\lambda_3$ increases between redshifts 1 and 0.5 and is significantly
stronger for the lower mass sample. Finally, we find that at all redshifts a
large $\lambda_3$ is strongly correlated with a small growth $G(z)$ and vice 
versa.

The correlations with $\lambda_3$ can now be compared with the dependence of
assembly on environment density. At all redshifts above zero, the correlation
coefficients between the overdensity in spheres of $2\,h^{-1}$Mpc and the three
measures of mass assembly are significantly weaker than those with $\lambda_3$.
Most remarkably, the correlation between $\delta_2$ and $z_{\rm form}$ is
strongest at $z\sim1$. However, at these times $z_{\rm form}$ correlates
significantly stronger with $\lambda_3$ than with $\delta_2$ indicating that the
density itself is not the primary cause for the suppression. As expected, 
the correlation between density and $\lambda_3$ grows strongly with time.

\begin{table}
\begin{center}
\begin{tabular}{llll}
\hline
\multicolumn{1}{c}  {}  & $z=0$ 	& $z=0.5$ 	& $z=1$  \\
\hline
\multicolumn{4}{c}{$M(0)=2-4\times10^{10}\,h^{-1}{\rm M}_\odot$}\\
  $Z(\lambda_3,\delta_2,z_{\rm form})$  & -2.81 & 3.71 & 6.74 \\
  $Z(\lambda_3,\delta_2,A)$             &  2.90 & -6.52 & -5.03 \\
  $Z(\lambda_3,\delta_2,G)$             &  -    & -10.66 & -9.24 \\
\hline
\hline
\multicolumn{4}{c}{$M(0)=2-4\times10^{11}\,h^{-1}{\rm M}_\odot$}\\
  $Z(\lambda_3,\delta_2,z_{\rm form})$  & -4.79 & 3.65 & 5.76 \\
  $Z(\lambda_3,\delta_2,A)$             &  3.98 & -4.08 & -0.61 \\
  $Z(\lambda_3,\delta_2,G)$             &  - & -8.78 & -10.57 \\
\hline
\end{tabular}
\end{center}
\caption{\label{tab:zscores}Z-scores $Z(x,y,z)$ reflecting the significance
in units of $\sigma$ that $\rho_s(x,z)>\rho_s(y,z)$ for the correlation coefficients 
given in Tables \ref{tab:corr_lambda} and \ref{tab:corr_lambda2}. Negative
values indicate the score for the alternative hypothesis $\rho_s(x,z)<\rho_s(y,z)$.}
\end{table}

The significance of the quoted correlations can be assessed by testing the
hypothesis that $\rho_s(x,z)>\rho_s(y,z)$ for three quantities $x,y,z$ 
respecting the fact that also $x$ and $y$ are correlated. This can be calculated 
using the method of \cite{Meng92}\footnote{ The original \cite{Meng92} test applies for 
Pearson correlation coefficients but can be easily adopted to rank correlations
by replacing all correlation coefficients and multiplying the standard error
of the z-transformed values with $1.03$ \citep[this is valid for samples larger
than 10 and $\rho_s<0.9$, see][]{Zar2007}.}.
The Z-scores (significance in units of $\sigma$) for the correlations in Tables 
\ref{tab:corr_lambda} and \ref{tab:corr_lambda2} 
are given in Table \ref{tab:zscores}. At redshifts 
$0.5$ and $1$, for all quoted correlation coefficients between $A$, $G$ and 
$z_{\rm form}$ with $\lambda_3$ and $\delta_2$, the differences are significant 
at more than $3\sigma$.

We will now briefly discuss the influence of numerical resolution effects in the
simulation on these results. Since the mass range of haloes shown in Table
\ref{tab:corr_lambda2} is resolved with 8 times more particles
in the $45\,h^{-1}{\rm Mpc}$ box,
albeit for a smaller sample, we compared all correlation coefficients
in the two simulations and found no significant deviations except
for the fit residuals $|\boldsymbol{\sigma}|$. At all redshifts,
correlations of assembly with $|\boldsymbol{\sigma}|$ were weaker and those with
$\lambda_3$ slightly stronger in the simulation with higher mass resolution.
The observation that the fit residuals depend on resolution is indicative of 
a minor resolution dependence in determining the tensor
eigenvalues from the flow that results in a slight weakening of correlations
with shear. Note also that the quoted results do not change significantly if haloes are
selected in a fixed mass range at some redshift $z>0$ instead of at $z=0$ except
that all correlations with $|\boldsymbol{\sigma}|$ are significantly weaker at
redshifts $z>0$ again reflecting a resolution dependence of the goodness of our
CSA model fit to the flow.

\subsection{Clustering behaviour of haloes in sheared regions}
\begin{figure}
\begin{center}
	\includegraphics[width=0.45\textwidth]{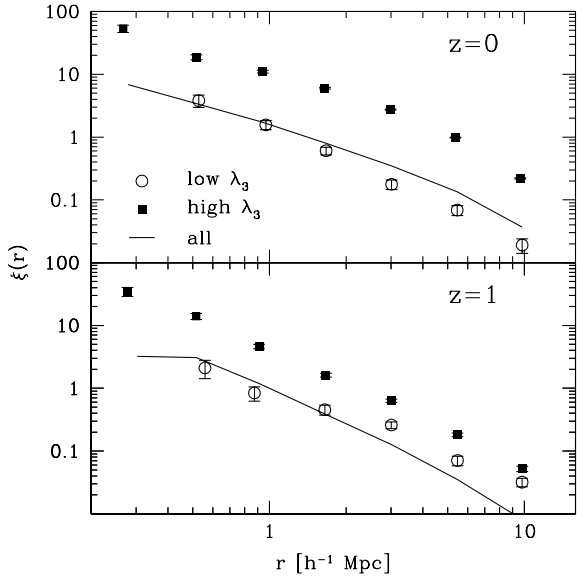}
\end{center}
\caption{\label{fig:corr_lam3}Two-point correlation functions for haloes with
masses between $2$ and $4\times10^{10}$ at redshift zero. The two panels
correspond to these haloes at redshift zero (top) and their main progenitor
haloes at redshift one (bottom). The samples have been split into those haloes
at the respective redshift with the 20 percent highest/lowest values in
$\lambda_3$, the largest eigenvalue of the rate-of-strain tensor. Tidally
stripped haloes are not included. Haloes experiencing strong tides, as probed by
$\lambda_3$, are much stronger clustered than haloes in low shear regions.}
\end{figure}
Tidal fields are strongest in the vicinity of massive haloes where also the
clustering of smaller haloes is substantially enhanced. In Figure
\ref{fig:corr_lam3} we show the correlation functions for haloes with the
largest/smallest eigenvalue $\lambda_3$ of the rate-of-strain tensor at
redshifts zero and their main progenitor haloes at redshift one. As expected,
haloes at redshift zero with high $\lambda_3$ are significantly more clustered
than average haloes while those with low $\lambda_3$ are less clustered. At
redshift one the situation is slightly different: only haloes with highest
$\lambda_3$ are significantly more clustered than average haloes. Furthermore,
haloes with the lowest $\lambda_3$ show a strong attenuation of the correlation
function at distances below $\sim 1\,h^{-1}\mbox{Mpc}$. This is a clear
indication that these haloes are only found in more underdense
regions. See also
the discussion in Appendix \ref{sec:lagrangian_assembly_bias} for the dependence
of clustering amplitudes on environmental overdensity in the initial density
field.

The tidal field, as quantified by the sheared flow around a halo, thus
shows a strong correlation with all quantities of mass assembly as well
as the dependence on clustering amplitude that is necessary to account
for the halo assembly bias. 

\subsection{Tidal stripping of halo mass}
\label{sec:sim_stripping}

A halo is prone to tidal mass loss once its Hill radius $r_{\rm hill}$,
defined in eq. (\ref{eq:hillradius}), with respect
to a massive neighbour becomes smaller than its virial radius $r_{\rm vir}$.
Defining haloes as objects with a fixed overdensity $\Delta_{\rm vir}$, we have
$m\propto \Delta_{\rm vir} r_{\rm vir}^3$ for the smaller
halo and $M\propto \Delta_{\rm vir} R_{\rm vir}^3$ for the larger halo. Then the condition
that $r_{\rm hill}<r_{\rm vir}$ is equivalent to
\begin{equation}
d < 3^{1/3} R_{\rm vir} \simeq 1.5 R_{\rm vir}.
\end{equation}
Thus, a halo is likely to suffer tidal mass loss once it has approached a more
massive halo to within 1.5 virial radii, assuming that the restricted three-body
approximation is justified and that the small halo remains nearly spherical.
Haloes that tidally loose mass to another halo should be considered as the most
extreme case of tidally suppressed mass growth: the range of their gravitational
influence has become smaller than their virial radius.

We will now quantify the population of these extreme examples of tidally
truncated mass assembly. As described in Section \ref{sec:massloss}, the
stripping implies an underestimation of halo mass at $z=0$ relative to the
assembly history at higher $z$, which is equivalent to an overestimation of the
formation redshift. It is thus important for our analysis to quantify the
abundance of these objects.

\begin{figure}
\begin{center}
	\includegraphics[width=0.45\textwidth]{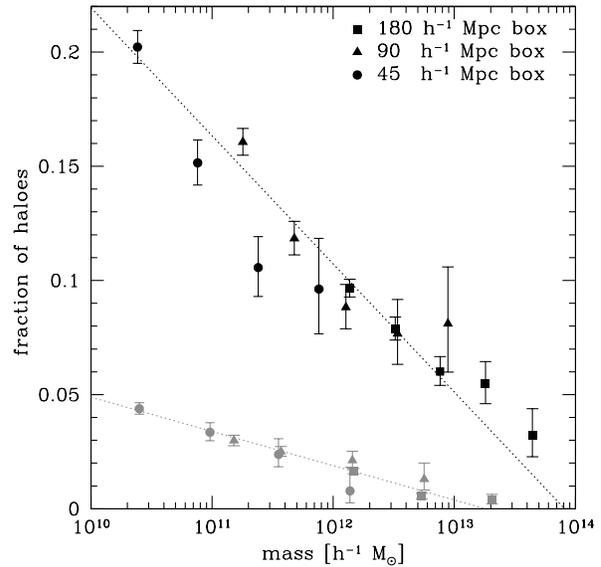}
\end{center}
\caption{\label{fig:massloss_fraction}Fraction of haloes that have $r_{\rm
Hill}<r_{\rm vir}$ at some point during their evolution (black symbols) and
fraction of haloes loosing at least 3 per cent of their mass to a more massive
halo (grey symbols) as a function of halo mass at redshift z=0. Errorbars
indicate the $1\sigma$ Poisson uncertainties. Results are shown for the three
simulations of different resolution described in the text. Dotted lines show
exponential fits to the data points.}
\end{figure}
 
In Figure \ref{fig:massloss_fraction} we show in grey the fraction of well
resolved haloes in our simulations that are found as isolated haloes at $z=0$
and lost at least 3 per cent of their particles to a more massive halo (Note
that this definition entails haloes that are actually detected as subhaloes at
$z>0$). At a given mass, this fraction is on the order of a few per cent. The
fraction is higher for smaller halo masses and reaches a value of $\sim4.5$ per
cent for $M\simeq2\times10^{10}\,h^{-1} M_\odot$.

However, these haloes do not represent the entire population of haloes that
underwent mass-loss in an encounter with a more massive halo. First, the time
resolution of our saved simulation snapshots does not permit to detect all
haloes that were at some point subhaloes of another halo, and second, it is not
guaranteed that the tidally unbound mass will be part of the more massive halo
after stripping. In order to get an upper limit on their abundance, we consider
now all those haloes for which the Hill radius is smaller than the virial radius
at some point during their evolution and that are thus prone to tidal mass-loss.
The fraction of haloes fulfilling this criterion is on the order of 20 per cent
at the lowest masses we considered (see Figure \ref{fig:massloss_fraction},
black line and symbols) in
agreement with the results of \cite{Diemand2007a} and \cite{Dalal2008}. The
lower number of those haloes for which actual mass-loss is detected implies that
the tidally unbound mass typically does not end up in the bigger halo. 

In Figure \ref{fig:harker_massloss} (left panel) we showed how the correlation
of formation redshift with overdensity $\delta_{2-5}$ changes when the haloes
that lost mass due to an encounter with a more massive object are removed from
the sample. The rank correlation decreases from $\rho_s(\delta_{2-5},z_{\rm
form})=0.18$ for the entire sample to $0.12$ after exclusion of the stripped
subset. Removing also the larger sample of those haloes that had $r_{\rm
Hill}<r_{\rm vir}$ during their evolution results in a slight increase of the
correlation coefficient to $0.16$ (all numbers are slightly higher when using
$\delta_2$ instead of $\delta_{2-5}$). Note that the criterion $r_{\rm
Hill}<r_{\rm vir}$ is roughly equivalent to selecting those haloes that were
within $1.5$ virial radii of another halo. Our results thus suggest that tidally
stripped haloes are only the most extreme cases of tidally truncated mass
assembly.

Furthermore, haloes that are at the apocenter of their orbit after passing
through a massive halo are also likely to have very large relative velocities
with respect to their environment. Considering the relative velocity between a
halo and the surrounding region within $1$ and $4\,r_{\rm vir}$ (cf. Section
\ref{sec:shearflow}), we find that haloes identified as stripped haloes indeed
show large relative velocities. These relative velocities will lead to extreme
velocity dispersions around haloes and are possibly dominating the results of
\cite{wang06} (their Figure 8, where median relative velocities are much larger
than the halo virial velocity only in the tail of the formation redshift
distribution). When removing the stripped haloes, we find that the correlation
between relative velocity and formation redshift weakens substantially.

%%%%%%%%%%%%%%%%%%%%%%%%%%%%%%%%%%%%%%%%%%%%%%%%%%%%%%%%%%%%%%%%%%%%%%%%%%%%%%%%%%%%%%
%%%% SECTION: DISCUSSION %%%%%%%%%%%%%%%%%%%%%%%%%%%%%%%%%%%%%%%%%%%%%%%%%%%%%%%%%%%%%

\section{Discussion}
The formation redshift of dark-matter haloes is highest for haloes that have the
lowest assembly rate at late times (cf. Section \ref{sec:assembly_rate}). Tying
this suppression of late growth to a physical effect that is most effective in
dense/strongly clustered regions can thus potentially explain the correlation 
of halo clustering and formation redshift, the so-called ``assembly bias''.

Tidal forces induce a sheared flow in the vicinity of small haloes that flow
along filaments of the cosmic web (Section \ref{sec:proximity}). To quantify
this effect, we measured the local influence of the larger scale tidal field by
considering the rate-of-strain tensor of the flow in the immediate vicinity of
haloes. 
We found that the progenitors of haloes of a given mass at $z=0$ that
have the lowest mass assembly rates experience significantly stronger shear 
during their assembly history than those haloes with high mass assembly
rates.

Tidal effects are expected to be strongest close to massive haloes. Whenever a
mass $M$ at distance $r$ from a smaller halo completely dominates the potential,
the tide in the direction of this mass is given by $2GM/r^3$, while the
perpendicular tides are given by $-GM/r^3$. In the rest-frame of the smaller
halo, matter is receding along the direction to the massive halo. Furthermore,
upon approaching the massive halo, the environment density will grow
substantially and as a consequence the correlation between density and shear
grows with time (see Section \ref{sec:TidalAssembly}). Since the abundance of
small haloes is enhanced near massive haloes, the population of tidally
suppressed haloes will be biased such that a dependence of formation redshift on
clustering amplitude and environment density arises as a secondary effect. The
more fundamental character of tides is reflected in the stronger correlation of
measures of halo growth with $\lambda_3$ than with environment density (Section
\ref{sec:TidalAssembly}).

The tidal influence of a larger halo on the accretion properties of neighbouring
smaller haloes can be readily understood from the restricted three-body-problem.
Stable orbits around the smaller halo exist only within its Hill sphere (see
Section \ref{sec:rhill}). A halo can only accrete particles (and other smaller
haloes) in the presence of a massive halo if they are contained in the Hill
sphere. The radius of the Hill sphere shrinks linearly with the distance to the
massive neighbour so that the amount of matter available for accretion is
becoming increasingly smaller. In addition, a halo is prone to mass loss when
the Hill radius becomes smaller than its virial radius. As shown in Section
\ref{sec:sim_stripping}, this happens roughly at a distance of $1.5$ virial
radii of the larger halo.

In the most extreme cases, small haloes can undergo very strong tidal mass-loss
when passing through the massive halo. A minor fraction of haloes of a given
mass are indeed not typical isolated haloes since they are once again found
outside the virial radius of a massive halo after their first pericenter passage
through its potential well. These haloes have also been observed in
resimulations of single massive haloes \citep{Gill2005,Diemand2007a,Ludlow2008}. 

While tidal mass-loss inside the virial radius of another halo is the most
violent and non-linear example, tidal suppression of halo growth seems to be 
already present in the quasi-linear regime. \cite{Keselman2007} recover the 
assembly bias also using the punctuated Zel'dovich approximation where the mass 
of haloes grows monotonously by definition and tidal mass-loss is thus impossible.

We explore the link between assembly bias and the initial linear density field
in Appendix \ref{sec:initial} where we compare the masses of haloes measured
from the simulations with the masses they should reach in an isolated
environment according to the EPS model. Our results indicate that the EPS
predicted masses are roughly a factor of 2-3 higher for the oldest haloes than
for the youngest. Furthermore, we find a significant correlation between this
mismatch and both the strength of the shear and large scale density at $z=0$ in
the simulation. This provides further evidence that indeed the environment, that
is not considered when determining the EPS-like masses, prevents the mass to
grow to the predicted value. Hence, while EPS would predict that the halo will
assemble mass, the strong shear that EPS does not account for prevents this
\citep[see also][]{Weygaert1994}. If the correct final mass is known, it can be
shown that the density dependence of formation redshifts can be reproduced using
a simple peak-background-split model (see Appendix \ref{sec:initial}).

A question that is opened by our analysis is whether the last major merger of
small haloes occurs preferentially perpendicular to their host filament such
that the mechanism we described is potentially also related to alignments
between halo spins and shapes and the surrounding large-scale structure
\citep[e.g][]{Dekel1985}. For small galactic dark matter haloes such strong
alignments between halo shape and the direction to the nearby cluster have been
found by \cite{Pereira2008} extending out to many virial radii of the cluster
and by alignments of both halo shape and spin with host filaments and sheets by
\cite{Hahn2007b} and \cite{Aragon06} \citep[see also][]{Basilakos06,Ragone2007}.
\cite{Lee2007} claim observational evidence for an alignment
of galaxy spins with the reconstructed tidal field. In the picture outlined in
this paper, the decreased probability for mergers or accretion along the
filament would preserve the spin vector alignment with the filament/sheet which
is established when merging occurs preferentially perpendicular to the
filament/sheet. 

We focused in this paper on the role of tidal effects in the assembly of haloes 
in different environments. We note that other global properties of haloes,
such as the substructure in them and their spin parameter, also correlate with
environment \citep[see e.g.]{Wechsler05,Gao06, Bett06}. The possible relation 
between these properties and the suppression of accretion onto haloes should be 
explored.

%%%%%%%%%%%%%%%%%%%%%%%%%%%%%%%%%%%%%%%%%%%%%%%%%%%%%%%%%%%%%%%%%%%%%%%%%%%%%%%%%%%%%%
%%%% SECTION: SUMMARY %%%%%%%%%%%%%%%%%%%%%%%%%%%%%%%%%%%%%%%%%%%%%%%%%%%%%%%%%%%%%%%%
\section{Summary and Conclusions}
Using a series of cosmological $N$-body simulations, we demonstrated that the
mass assembly of haloes is influenced by tidal effects in the
vicinity of massive haloes. The resulting suppression of mass growth at late
times is reflected in an increase in formation redshift. The space density of
small haloes near massive haloes is enhanced (since their spatial
cross-correlation is high and positive). Thus, as a secondary effect, a tidally
suppressed assembly of halo mass directly translates into biased spatial
distributions of young and old haloes identified at $z=0$: old haloes will be
found more likely in regions of high density than younger haloes.

The presence of tides has an important effect on the infall behaviour onto a 
halo. To quantify this, we measured the influence of external tides by considering 
the rate-of-strain tensor, and especially its largest eigenvalue $\lambda_3$, 
fitted to the flow between 1 and 4 virial radii around haloes. In detail, we 
found that for small haloes:
\begin{enumerate}
\item High formation redshift is related to a low assembly rate at late times.
Thus, any spatially biased process that suppresses mass growth at late times can
potentially lead to an age dependent bias of halo clustering.

\item Among all measures of environment that we considered, assembly rate at
redshifts $z>0$ correlates most strongly with the largest eigenvalue $\lambda_3$
of the rate-of-strain tensor, and especially so when measured at high redshift,
$z\sim1$. The correlation of assembly rate with the density of the environment
is weaker. We find that haloes with a low assembly rate have suppressed infall
or even outflow in the region between 1 and 4 virial radii along one direction.
Furthermore, we find that this direction corresponds, with little scatter,
to the direction to that neighbouring halo that has the strongest tidal influence.

\item The strain eigenvalue $\lambda_3$ (probing velocity shear) is strongly
correlated with the tidal field due to the
neighbouring halo that exerts the strongest tides. This correlation has
been verified using a restricted three-body approximation for the tidal field.
Thus, tidal forces induce a velocity shear in the flow around dark matter
haloes and can render accretion ineffective by altering the convergence
of the accretion flow onto a halo.

\item The correlation between $\lambda_3$ (probing velocity shear and tides) and
density grows with time. Thus, haloes in strongly sheared regions are much
more strongly clustered than average and associated with high density regions
onto which they fall.
Hence the suppression of mass growth is stronger in regions of high density.

\item Haloes can undergo mass loss due to the tidal influence of a larger halo
when their Hill radius becomes smaller than their virial radius. This already
happens at a distance of $1.5$ virial radii of the larger halo and will
be most severe inside its virial radius. Roughly 20 per
cent of the $2-4\times10^{10}\,h^{-1}{\rm M}_\odot$ fulfill the condition for
mass loss at some point of their history. These haloes are the most extreme
cases of tidal suppression of assembly. 

\item The mismatch between the mass that the Extended Press-Schechter formalism 
would attribute to a halo and the mass that is measured for it in the simulation
correlates with formation time and with the eigenvalue $\lambda_3$ indicating
a dependence of this mismatch on tides (this is shown in the appendix).

\end{enumerate}

From the evidence presented above, we conclude that the tidal field due to
neighbouring massive haloes exerts a shear on the flow surrounding small haloes.
In the rest-frame of the small halo, this is an outward flow along the direction
to the more massive halo, tidally suppressing the growth of the small halo. This
is especially effective in filaments since most of the matter is located along
the direction from which it is least effective for the halo to accrete.

The proximity to more massive haloes, where the clustering is enhanced, 
leads to a spatial bias of haloes that experience strong tides,
especially haloes below the non-linear mass, and thus also to a bias of 
the population of ``old'' and ``young'' haloes of the same mass. We conclude 
that this is an important factor in the emergence of the {\it assembly bias}.

\section*{Acknowledgements}
OH acknowledges support from the Swiss National Science Foundation. CP thanks
the participants of the Aspen workshop ``Modelling Galaxy Clustering'' for
discussions. AD's research has been supported by an ISF grant, by a GIF grant
I-895-207.7/2005, by a DIP grant, by the Einstein Center at HU, by NASA ATP 
NAG5-8218 at UCSC, and by a French-Israel Teamwork in Sciences. All simulations 
were performed on the Gonzales cluster at ETH Z\"urich, Switzerland.

%%%% APPENDIX %%%%%%%%%%%%%%%%%%%%%%%%%%%%%%%%%%%%%%%%%%%%%%%%%%%%%%%%%%%%%%%%%%
\appendix
\section{Assembly Bias in the Initial Density Field}
\label{sec:initial}
\subsection{Predicting the assembly bias from the initial density field}
Can we predict the amplitude of the assembly bias directly from
the linear density field?
This requires being able to forecast the formation redshift, final mass
and local Eulerian density of a halo directly from the initial conditions.

\subsubsection{Halo formation redshift} 
\begin{figure}
\begin{center}
    \includegraphics[width=0.45\textwidth]{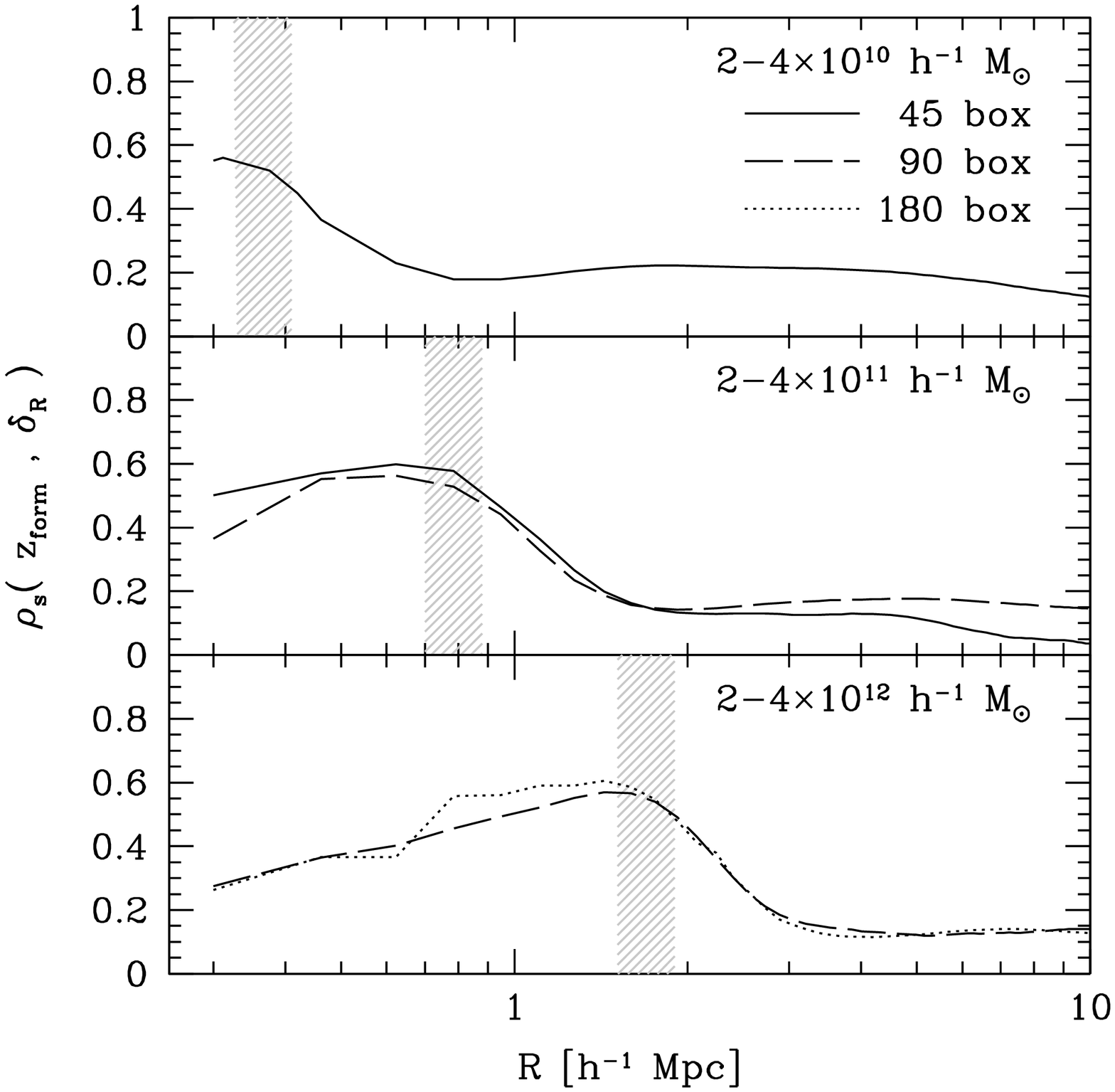}
\end{center}
\caption{\label{fig:corr_scale} Spearman rank correlation coefficient $\rho_s$
between halo formation redshift $z_{\rm form}$ and ``peak height'' of the
Lagrangian density field smoothed with a real space top hat filter on scale $R$
as a function of the filter scale. Results are shown for three mass ranges.
Results for the different simulations are indicated by different line styles.
The shaded regions indicate the half-mass range $M/2$ for the corresponding mass
bins. This analysis reveals: (1) The correlation coefficient peaks as expected
for haloes of mass $M$ on scales around $M/2$; and (2) the correlation with the
large scale density field on several $h^{-1}{\rm Mpc}$ is significant.}
\end{figure}

What determines the formation redshift
of a halo? In Figure \ref{fig:corr_scale}, we show the Spearman rank correlation
coefficient between the formation redshift $z_{\rm form}$ and the linear
overdensity field evaluated at the centre of mass of the Lagrangian patch that
will form a halo at $z=0$. The overdensity field is computed after smoothing the
density field with a top-hat filter of radius $R$. The rank correlation
coefficients with $z_{\rm form}$ are plotted against $R$.
Our results show that the formation time of a halo of mass $M$ is most strongly
determined by the value of the linear overdensity smoothed on a scale $\sim
M/2$. This is what one trivially expects for spherical perturbations.
Tides on slightly larger scales (but still internal to the protohalo) also
correlate with $z_{\rm form}$ but to a lesser degree (not shown in the figure).
The tidal suppression of halo assembly discussed in the previous sections
manifests itself in the initial conditions as a residual correlation between
$z_{\rm form}$ and $\delta_R$ for large smoothing scales. The rank correlation
coefficient is as high as $0.2$ even for smoothing radii of a few Mpc. The
presence of nearby density fluctuations influences the formation time of a given
halo. 

\subsubsection{The assembly bias}

If the final halo masses are known, the assembly bias can be perfectly
reproduced starting from the initial density field and using a simple
peak-background split model where small-scale fluctuations are responsible for
halo collapse and large-scale fluctuations determine the local density of the
halo environment.

Inspired by Figure \ref{fig:corr_scale}, we assume that the formation
redshift
$\hat{z}_{\rm form}$ of a halo of mass $M$ is fully determined by
the ``peak height'' $\delta_{M/2}$.
Denoting by $D_+(z)$ the linear growth factor of density fluctuations, we thus
solve the condition for collapse of a spherical density perturbation
\begin{equation}
\frac{ D_+(\hat{z}_{\rm form})}{D_+(z_i)}\,\delta_{M/2}=\delta_{c},
\end{equation}
where $z_i$ is the initial redshift. We find that a value of $\delta_{c}=2.3$
leads to the best agreement between $\hat{z}_{\rm form}$ and $z_{\rm form}$
obtained from the $N$-body simulation.

In a second step, we determine the evolution of the background density field
$\delta_5$ on scales of $R_b=5\,h^{-1}{\rm Mpc}$. Small haloes that survive
until $z=0$ will preferentially form in regions that are underdense on these
scales in the initial density field. They will then migrate into regions of
higher density over time. We determine the evolution of the background density
using first-order Lagrangian perturbation theory. Given the eigenvalues
$\Lambda_i$ of the initial tidal field on scales of $R_{b,0}$, the evolution of
$\delta_5$ at the positions of the haloes is given before shell-crossing by
\begin{equation}
\delta_{5}(z) = \frac{1}{\prod_{i=1}^{3}
\left[1-D_+(z)\,\Lambda_i\right]}-1
\end{equation}
in the Zel'dovich approximation \citep{Zeldovich70}.

\begin{figure}
\begin{center}
    \includegraphics[width=0.45\textwidth]{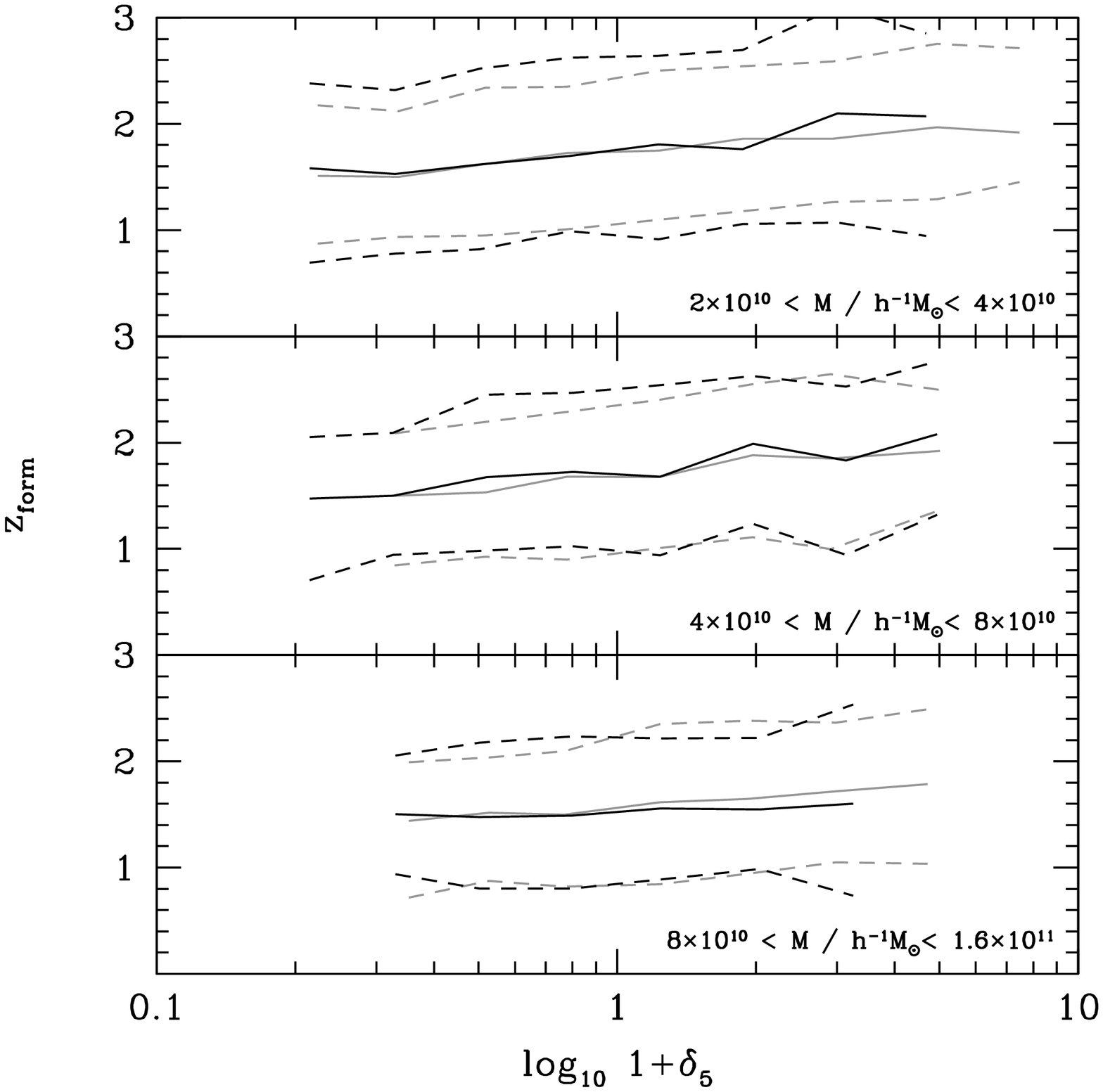}
\end{center}
\caption{\label{fig:harker_eps} Halo formation time as a function of
environmental density on scales $5\,h^{-1}{\rm Mpc}$ obtained from the $N$-body
simulations ({\it grey}) and obtained from the initial conditions data using a
combination of two spherical collapse models for the evolution of the background
and the peak ({\it black}), the mass of the halo is taken from the $N$-body
simulation. Stripped haloes are not considered. Solid lines indicate the median,
dashed lines the $\pm1\sigma$ spread around the median.}
\end{figure}

For each halo in the $N$-body simulation, we thus obtain an environmental
density at $z=0$ and the redshift of collapse $\hat{z}_{\rm form}$ by knowing a
priori its final mass and its Lagrangian position. In Figure
\ref{fig:harker_eps} we show the median formation redshift $\hat{z}_{\rm form}$
(in grey) as a function of density $\delta_5$ for three mass bins.\footnote{We
have checked that results are not influenced by the small fraction of haloes
that already experienced shell-crossing on the $5\,h^{-1}{\rm Mpc}$ scale.} The
corresponding relation is shown also for the quantities directly obtained from
the $N$-body simulation (in black). We find very good agreement with the
$N$-body data suggesting that this simple model suffices to predict the
correlation between formation redshift and environment seen in the simulations
if the correct final mass is known.

\begin{figure}
\begin{center}
    \includegraphics[width=0.45\textwidth]{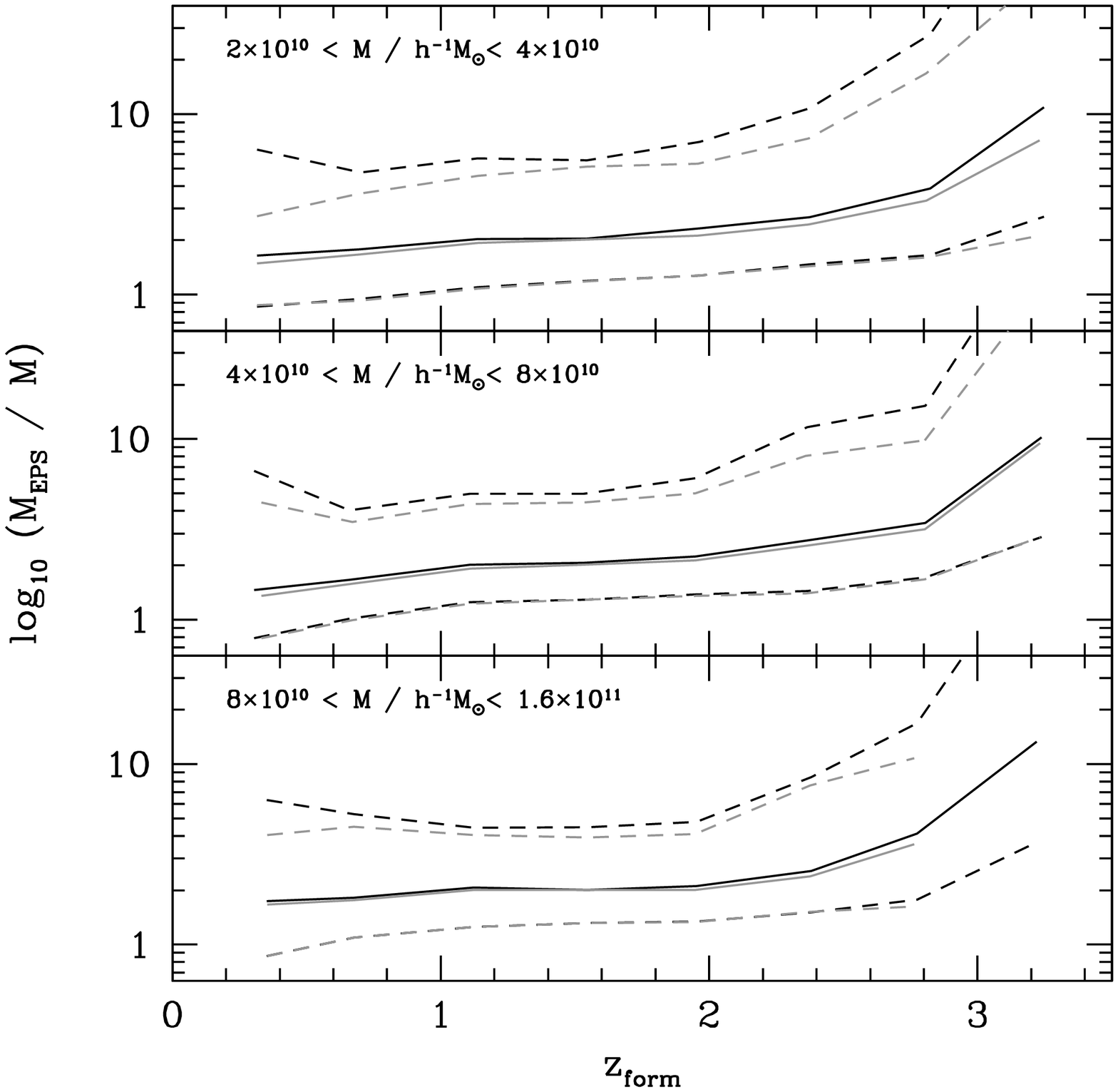}
\end{center}
\caption{\label{fig:epsmasses} The ratio of the mass determined in the initial
conditions using the EPS model to the mass measured in the $N$-body 
simulation as a function of halo
formation time. Solid lines indicate the median, dashed lines the 16th and 84th
percentiles. Shown is the relation for haloes that are not undergoing tidal
mass-loss (grey) and for all haloes (black) in three mass bins.}
\end{figure}

Replacing the actual halo masses with those predicted at the Lagrangian halo
locations by the Extended Press-Schechter (EPS) formalism is sufficient to
completely erase any dependence of the formation redshift on environmental
density. 
The spherical collapse model (on which the EPS method is based) tends
to overestimate halo masses, particularly for those haloes that are located
within strongly sheared flows.
In Figure \ref{fig:epsmasses} we show how the formation redshift $z_{\rm form}$
depends on the ratio between the mass $M_{\rm EPS}$ and the mass $M$, where
$z_{\rm form}$ and $M$ are determined from the $N$-body simulations. We find
that the mismatch increases from a median ratio of about 1.5 at low formation
redshifts to about 3 at high formation redshifts for the considered masses
\citep[cf. Figure 3 in][]{wang06}. When we exclude the mass-stripped haloes
(grey lines) the scatter is significantly reduced.
The ratio between the EPS-predicted mass and the mass found in the $N$-body
simulations correlates with the largest eigenvalue $\lambda_3$ of the
rate-of-strain tensor as strongly as with the halo formation redshift. For
haloes at $z=0$ with masses between 2 and $4\times10^{10}h^{-1}{\rm M}_\odot$
the Spearman rank correlation coefficient between $M_{\rm EPS}/M$ and the shear
eigenvalue $\lambda_3$ is 0.29. It is 0.27 between $M_{\rm EPS}/M$ and $z_{\rm
form}$. This provides further evidence that tidally suppressed accretion is
responsible for the existence of the assembly bias. The proximity to a massive
halo is reflected in the correlation coefficient with density obtained with the
top-hat filter at z=0. We find that the correlation coefficient with $M_{\rm
EPS}/M$ is 0.22 for density determined on scales of $5\,h^{-1}{\rm Mpc}$, 0.31
for $2\,h^{-1}{\rm Mpc}$ and 0.41 for $0.5\,h^{-1}{\rm Mpc}$. 

\subsection{Lagrangian assembly bias}
\label{sec:lagrangian_assembly_bias}
\begin{figure}
\begin{center}
    \includegraphics[width=0.45\textwidth]{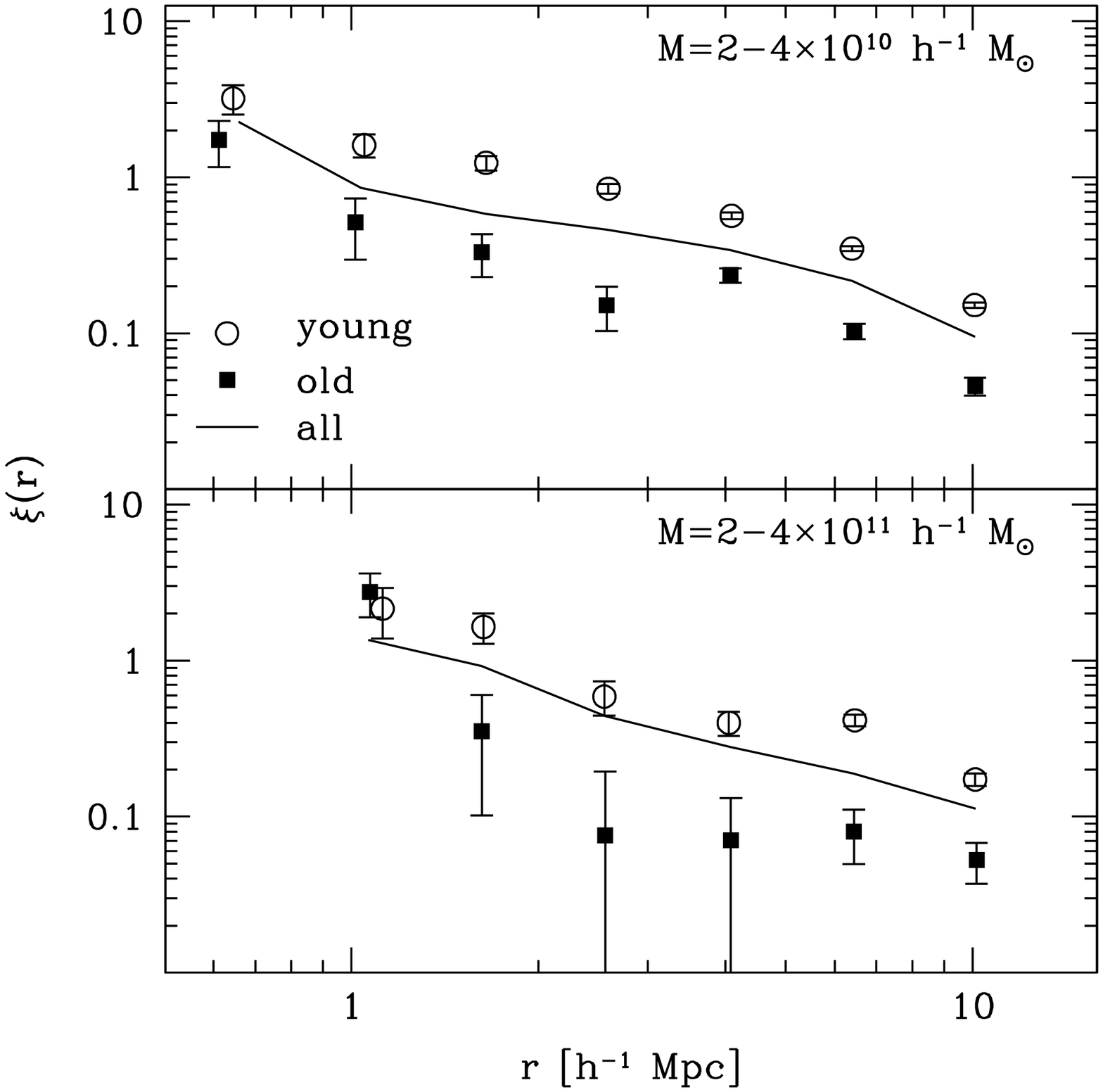}
\end{center}
\caption{\label{fig:corrfun} Two-point correlation functions of the Lagrangian
centres of mass in the initial density field ($z=78.5$) for haloes with masses
$2-4\times10^{10}\,h^{-1}{\rm M}_\odot$ at $z=0$ (upper panel) and
$2-4\times10^{11}\,h^{-1}{\rm M}_\odot$ (lower panel). The solid line shows the
correlation function for all haloes, while open circles and filled squares
correspond to the 10 per cent of these haloes with lowest and highest formation
redshifts, respectively.}
\end{figure}
It is interesting to verify whether the assembly bias is already present in the
spatial distribution of protohaloes in the linear density field. The Lagrangian
bias is given by $b_L=b_E-1$ if $b_E$ is the Eulerian bias
\citep[e.g.][]{MoWhite1996}. For $M<M_\ast$ haloes $b_E<1$. Thus, if $b_{E,{\rm
old}}>b_{E,{\rm young}}$ it follows that $0\geq b_{L,{\rm old}}>b_{E,{\rm
young}}$ and thus the old haloes should have a lower Lagrangian clustering
amplitude than the young ones.

To verify this, we compute the two-point correlation of the Lagrangian patches
of old and young haloes (selected at $z=0$) at the initial redshift of our
simulations.
Figure \ref{fig:corrfun} shows the correlation function of the protohaloes
that will form objects with masses between $2$ and
$4\times10^{10}\,h^{-1}{\rm M}_\odot$ at $z=0$.
The corresponding correlations at $z=0$ have been presented in Figure
\ref{fig:harker_massloss}. At $z=0$ the oldest haloes cluster more strongly than
the youngest ones. For the Lagrangian patches, as expected, the opposite is
true: the oldest (youngest) haloes are significantly less (more) clustered than
the total.
This indicates that, already in the linear density field, the environments
surrounding the Lagrangian patches of young and old haloes have different
properties.
In particular, we find that both young and old haloes at these masses form in
initially underdense (i.e. with a median $\delta_5<0$) regions. However, the
Lagrangian patches of young protohaloes reside (on average) in more underdense
regions which, in a Gaussian random field, are more strongly clustered.
Using the mean squared overdensities at the halo
centres as a proxy for their clustering amplitude (i.e. assuming a linear
bias model independent of halo age for the halo density),
$B=\left<\delta_5^2\right>^{1/2}$, we find
$B_{\rm old}/B_{\rm all}=0.8$ and $B_{\rm young}/B_{\rm all}=1.2$.
On the other hand, with time, the protohaloes will collapse and
flow towards denser regions thus changing their clustering properties.
At $z=0$ we find  $B_{\rm old}/B_{\rm all}=1.8$ and
$B_{\rm young}/B_{\rm all}=0.33$.
The reversal of the assembly bias is then a simple result of the evolution
of the halo environmental density. Older haloes have thus always resided
in more overdense regions than younger haloes.

\label{lastpage}
\end{document}